\documentclass[useAMS,usegraphicx]{mn2e}
\usepackage{wasysym,subfig}
\input epsf.tex

\title[Dynamical Erosion of a Moon forming disk]{ A dynamical context for the origin of Phobos and Deimos} 
\author[Hansen] {Bradley M. S. Hansen$^1$\thanks{E-mail:hansen\@astro.ucla.edu}\\
$^1$Mani L. Bhaumik Institute for Theoretical Physics, Department of Physics \& Astronomy, \\ University of California Los Angeles, Los Angeles, CA 90095}

\begin{document}

\date{submitted}


\maketitle

\label{firstpage}

\begin{abstract}

We show that a model in which Mars grows near Earth and Venus but is then
 scattered out of the terrestrial region yields a natural pathway to explain the low masses of the 
Martian moons Phobos \& Deimos. In this scenario, the last giant impact experienced by Mars is followed by
an extended period (tens to hundreds of Myr) of close passages by other planetary embryos.
These close passages perturb and dynamically heat any system of forming satellites left over
by the giant impact and can substantially reduce the mass in the satellite system (sometimes to zero).
 The close passage of massive perturbing bodies also offers the opportunity
to capture small objects by three-body scattering. Both mechanisms lead to low mass moon systems with
a substantially collisional history.

\end{abstract}


\section{Introduction}

The discovery of planets orbiting other stars has led to an ongoing revolution in our understanding
of the origins of planets and their system architectures. In addition to spawning theories about
the evolution of these distant systems, this has also led to new insights  regarding the formation of the
planets in our own Solar system. Perhaps the biggest change with regards to classical theories of
planet origin is the appreciation that the early evolution of a planetary system is far more dynamic than
originally believed, and that the current positions of planets need not indicate the original location
of the material from which they formed.

In the case of our own Solar system, we benefit from an amount of observational detail that is not
feasible to obtain in other systems. One aspect of this is that we can observe even relatively small satellites of the
planets in our Solar system, whose origins must therefore be incorporated into the formation scheme.
 Viewed in this light, the variation in the satellite systems of the Terrestrial planets is striking, given that
all the planets are believed to be broadly the result of the same collisional accumulation process (e.g. Stevenson, Harris \& Lunine 1986).
Mercury and Venus possess no known long-term moons, and the moons of Mars -- Phobos and Deimos -- are much
smaller ($\sim 10^{-3}$ in mass) than that of the Earth. Thus, it appears as though satellite formation does not
scale simply with planet mass.

Indeed, it was initially hypothesized that Phobos and Deimos were captured asteroids (Singer 1968; Smith 1970). 
This point of view  is
 supported by their irregular, cratered, shapes, low densities, and the similarity of their colours with those
of outer main belt asteroids (Christensen et al. 1977; Tolson et al. 1978;
Duxbury \& Veverka 1978; Pang et al. 1978; Pollack et al. 1978; Pajola et al. 2013; Fraeman et al. 2014). On the other hand, their orbits are almost circular, and
lie close to the equatorial mid-plane of the planet, suggesting that substantial dissipation is required
to produce the current dynamical configuration. Formation in an impact generated disk, like in the 
case of the Moon, would indeed provide sufficient dissipation to damp the resulting debris disk down
to such a configuration (Goldreich 1965), but then we are left with
the question of why the bodies are so much smaller than the Moon, and why do they share morphological
and photometric properties of asteroids?

This dichotomy between appearance and orbital parameters has shaped the subsequent discussion of the 
origins of Phobos and Deimos, without an obvious clear resolution. The success of a capture model will
depend on a plausible origin for the dissipation necessary to capture and circularise the orbits. 
Capture has been proposed via nebular or atmospheric drag (Hunten 1979; Pollack, Burns \& Tauber 1979) or
by the collision or disruption of asteroid binaries (Walker 1969; Pajola et al. 2013). The orbit of
Phobos can be reasonably circularised by tidal dissipation but it is difficult for a body as distant
as Deimos to become circularised by tides. Furthermore, generic capture models imply highly eccentric
orbits which would cause Phobos and Deimos to cross, and rapidly collide (Szeto 1983).

 Collision models, on the other hand, produce a disk of debris which condenses to form small 
bodies that eventually assemble to form satellites
in the equatorial plane. Most quantitative attempts to examine this scenario focus on impactors
associated with the largest observed craters on Mars, such as the Borealis basin (Craddock 1994, 2011;
Marinova, Aharanson \& Asphaug 2008; Rosenblatt 2011). Within this scenario, the masses of the Martian moons are smaller than the Terrestrial
moon because of a smaller impact velocity and consequently smaller disk. The precise mass of debris
is somewhat uncertain, depending on the details of the impact and the vapour content of the ejected
material (Craddock 2011; Rosenblatt \& Charnoz 2012; Citron, Genda \& Ida 2015). Indeed, some
(e.g. Stevenson 1987) have questioned whether it is even possible to form a disk, given that the
impact velocity needed to vaporise rock (e.g. Ahrens \& O'Keefe 1972) is comparable to the escape velocity from the surface of Mars.
 The most detailed
calculations (Hyodo et al. 2017) invoke impactors a few percent the mass of Mars, producing a debris disk whose mass
is a few percent that of the impactor. This is still orders of magnitude larger than the current
satellite masses, but it is suggested that much of that material fell into Mars on shorter timescales.
The angular momentum in such impacts also suggests that the bulk of the material goes into orbit at
or just inside the Roche limit, with a small fraction at larger radii.
 The radial distribution of mass is important in the face of geochemical calculations
of the crystallisation of the impact-generated disk into solids (Ronnet et al. 2016). If the material
passes through a liquid phase as occurs in traditional models of impact generated disks (Thompson \& Stevenson 1988;
Ward 2012), it is expected to show high concentrations of Olivine, which should yield observable
absorption bands in the 1--2$\mu m$ range. This is not observed, although evidence
for phyllosilicates is found in the thermal infrared (Giuranna et al. 2011). Ronnet et al. propose that direct
gas--solid condensation in a more extended, purely gaseous, disk (e.g. Rosenblatt \& Charnoz 2012) could yield small enough grains to
match the observations, but such extended disks are not favoured by the simulations of impacts (Citron et al. 2015, Hyodo et al. 2017).
This has led to revised scenarios in which the remaining satellites are formed from the outer edge of the
original disk and were shepherded outwards by larger moons that subsequently migrated inwards due to
tides and were 
disrupted (Rosenblatt et al. 2016; Hesselbrock \& Minton 2016).

To date, the discussion of moon formation in a Martian context has taken place within the traditional
planet formation framework, in which Mars condensed out of material more or less at its current location,
and has experienced little orbital evolution over its history. However, the in situ accretion of Mars
in traditional nebular disk models yields a problematically large mass relative to the observed value
(Wetherill 1986; Chambers 2001). Models in which the terrestrial planets are all initially seeded in a narrow
annulus inside~1 AU (Hansen 2009; Walsh et al. 2011; Brasser 2013; Walsh \& Levison 2016) can solve this problem by explaining Mars as a 
body that diffused outwards due to scattering off the larger planets and was thereby starved of
material to accrete. This is consistent with geochemical analyses which suggest that the accretion
of Mars finished early ($\sim \rm 10 Myr$ -- Nimmo \& Kleine 2007; Dauphas \& Pourmand 2011) relative to Earth ($\sim \rm 100 Myr$ -- Jacobson et al. 2014).
This has led to
a description of  Mars as a  ``surviving planetary embryo'', in the sense that it underwent far less collisional and
accretional evolution than the other terrestrial planets.

This model for the early evolution of Mars also offers a fresh perspective for the origin of the Martian moons. If the
moons are indeed the result of a giant impact, then the resultant moon forming disk is subjected to an
extended history of gravitational perturbations after formation, as the orbit continues to evolve by
close gravitational encounters with other bodies. Such encounters should then erode the disk over time, as
has also been suggested for families of giant planet irregular satellites (Li \& Christou 2017).
On the other hand, these gravitational encounters also offer
a new, and far more natural, opportunity for capture by virtue of direct three-body interactions during the close encounters
that cause Mars orbital evolution. A similar
process has been suggested for the formation of the irregular satellite populations of the giant
planets (Nesvorny, Vokrouhlicky \& Deienno 2014).

Our goal in this paper is to
 examine, in more detail, the orbital histories of Martian analogues in
the context of the annular birth scenario, and to see how this affects the properties
of a satellite population bound to Mars.
 In \S~\ref{History} we will examine the orbital histories of Mars analogues 
 and how this shapes the history
of gravitational perturbations for each system. In \S~\ref{Consequences} we then examine both
the erosion of possible moon forming disks and the capture of unbound material by three body
interactions. In \S~\ref{Aftermath} we examine the nature of the small body populations that
are produced in this manner, and estimate how they are shaped by collisional and tidal processes
to produce the population seen today.

\section{The Orbital Histories of possible Mars Analogues}
\label{History}

Whether it is by capture or collision, the origins of the Martian moons are associated
with encounters between the proto-Mars and other solar system bodies. Thus, any origin
scenario for the moons must take into account the orbital history of Mars and bodies
that pass near it. In this section, we will examine the orbital evolution of Mars-like
bodies within the annular origin model discussed in Hansen (2009).
This starts with $2 M_{\oplus}$ of material
in planetary embryos of mass $0.005 M_{\oplus}$, initially placed on circular, almost coplanar
orbits in an annulus spanning a limited range of semi-major axes.
 A Jupiter-mass planet at $5.2$AU serves to
restrict outwards diffusion by scattering. We consider two sets of annuli, in order to 
assess whether the annulus thickness has an influence on the results. We simulate
25 systems with an annulus from 0.7--1~AU (\#1 -- \#25), and 25 systems with annulus from 0.8--1~AU (\#26 -- \#50).
We also perform 10 additional, `high resolution' simulations, in which the same mass
is apportioned amongst bodies of mass $0.0017 M_{\oplus}$. Five of these start with
the wider annulus (HR1--HR5) and 5 with the narrower annulus (HR6--HR10).
At this higher resolution scale, the mass of the original embryos 
is the same as that inferred for the impactors that formed the largest basins on
Mars (e.g. Craddock 2011), and which are postulated to be responsible for the moon-forming disk.
 This makes for a total of 60 simulations.

We perform the simulations with the {\tt Mercury} code (Chambers 1999), which simulates
the direct gravitational interactions between the bodies in the system, and incorporates
the evolution through scattering and collision. Collisions are assumed to result in perfect
accretion. This is an oversimplification but has been shown to yield similar final results
in terms of the number and masses of final bodies as well as their bulk assembly timescales
(Kokubo \& Genda 2010). For the purposes of discussions in future sections, it is worth
noting that the principal effect of a more detailed collision model is the production
of a population of low mass debris (Stewart \& Leinhardt 2012) which can lengthen the
time to completely clear the system (Chambers 2013).

Each system is evolved for at least 100~Myr, with a 4~day timestep, in order to resolve
possible Mercury analogues.
 Most systems have reached a 
dynamically stable configuration by this point, but a few systems are still dynamically
coupled (i.e. orbits of two or more bodies overlap) at this age. The evolution of this subset of systems
is continued until further interactions result in a dynamically decoupled system.
Our goal in these simulations is to seek systems which contain a planet sufficiently close in its physical
and orbital properties to be considered a Theoretical Mars Analogue (TMA).

\begin{figure}
\includegraphics[width=84mm]{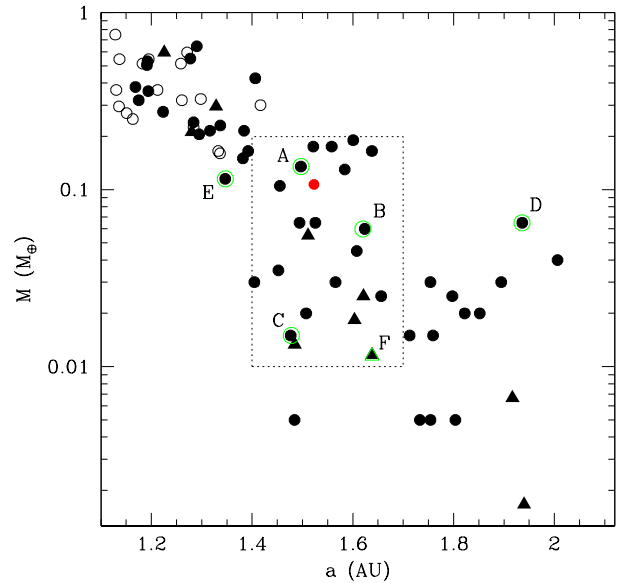}
\caption[MarsCases.ps]{The points show surviving planets after the simulated
terrestrial planets have evolved to the point of being dynamically decoupled.
Only planets with semi-major axis from 1.1--2.13~AU are shown. There is one additional
point at 3.04~AU and $0.03 M_{\oplus}$ off the right of this plot.
The filled points represent
those which are the outermost terrestrial planet in each system. Circles represent the results
of the standard runs and triangles show the results of the higher resolution runs. The dotted
rectangle represents our definition of likely TMA, as discussed
in the text. The red point indicates the observed mass and semi-major axis
of Mars. The points encircled in green are those referenced by the nearby alphabet symbols.
\label{MarsCases}}
\end{figure}

Figure~\ref{MarsCases} shows the distribution of mass and semi-major axis for the
outer parts of the resulting dynamically decoupled systems. The solid points represent
the properties of the outermost terrestrial planet in each system (i.e. not counting
Jupiter). In some cases, these planets are massive enough and close enough to 1~AU as
to be better considered an analogue of Earth or Venus rather than Mars. We define
such systems as not containing a TMA. For the rest, we define this to
be the sample of TMA (a total of 39/60 of the simulated systems contain a TMA). The quantitative limits we use for this are
a mass $M < 0.2 M_{\oplus}$ and a semi-major axis $a > 1.4$AU. We also wish to subdivide
this class further, defining  ``likely'' TMA by $0.01 M_{\oplus} < M < 0.2 M_{\oplus}$
and $1.4 < a < 1.7$~AU. This is to restrict ourselves to objects whose parameters are not too
dissimilar from those of the observed Mars, in case variations in such parameters prove to
be important in the forthcoming calculations. A total of 22/60 systems (37\%) contain a likely TMA.

\begin{table*}
\centering
\begin{minipage}{140mm}
\caption{Parameters of the outermost terrestrial planet from each simulation. This table
shows a representative sample. The full table is available online.
 \label{Outcomes}}
\begin{tabular}{@{}lccccccc@{}}
\hline
 $\#$ &
Mass & Semi-Major Axis & Eccentricity & Inclination & Last Impact & Decoupling & TMA? \\
    &($M_{\oplus}$) & (AU) &  &  ($^{\circ}$) & (Myr) &  &  \\
\hline
1 & 0.242 & 1.284 & 0.123 & 6.27 & 41.5 & HI  & No \\
2 & 0.005 & 1.484 & 0.077 & 29.9  & $\cdots$ & HI & P \\
3 & 0.115 & 1.347 & 0.125 & 1.91 & 104.7 & HO  & No  \\
4 & 0.505 & 1.191 & 0.028 & 3.75 & 30.4 & HI & No \\
5 & 0.015 & 1.487 & 0.195 & 6.42 & 0.2 & HI & L \\
\hline
\end{tabular}
\end{minipage}
\end{table*}

This stricter definition does leave behind some possible alternative TMA. We designate these
as 'P', for Possible. They can be of similar mass to our TMA, but with larger semi-major axis,
or they can fall below the mass limit we impose. In some of these cases, 
 the outermost planet is a true planetary embryo in the sense that it is one of the original seed 
particles that survives to the end of the simulation without having undergone a collision. It is sometimes hypothesized that
this might indeed be the case for the observed Mars (e.g. Dauphas \& Pourmand 2011) but the original embryo mass in
our simulations is $0.005 M_{\oplus}$ or less, and quite a bit smaller than the observed Mars mass. 

\subsection{Orbital Histories}

The presence of a TMA within the region we have defined implies
that the planet must have both grown by accretion (because we require true
analogues to be more massive than the initial embryos) and have migrated outwards from the initial annulus.
We wish to understand the relative importance and timelines of these two processes (accretion and migration)
in the creation of
a TMA.

\begin{figure*}
\centering
    \subfloat[]{%
      \includegraphics[width=.3\textwidth]{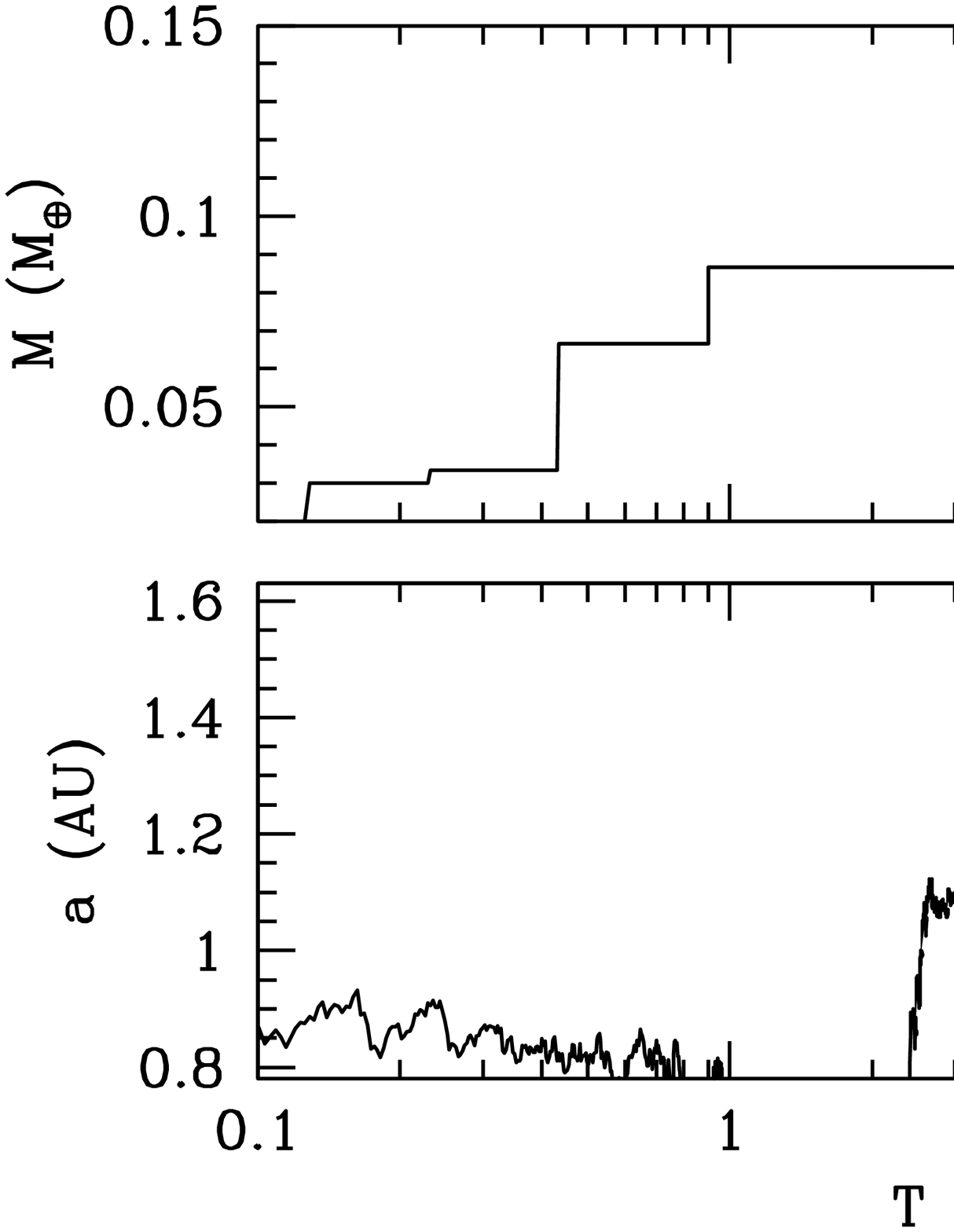}}\hfill
    \subfloat[]{%
     \includegraphics[width=.3\textwidth]{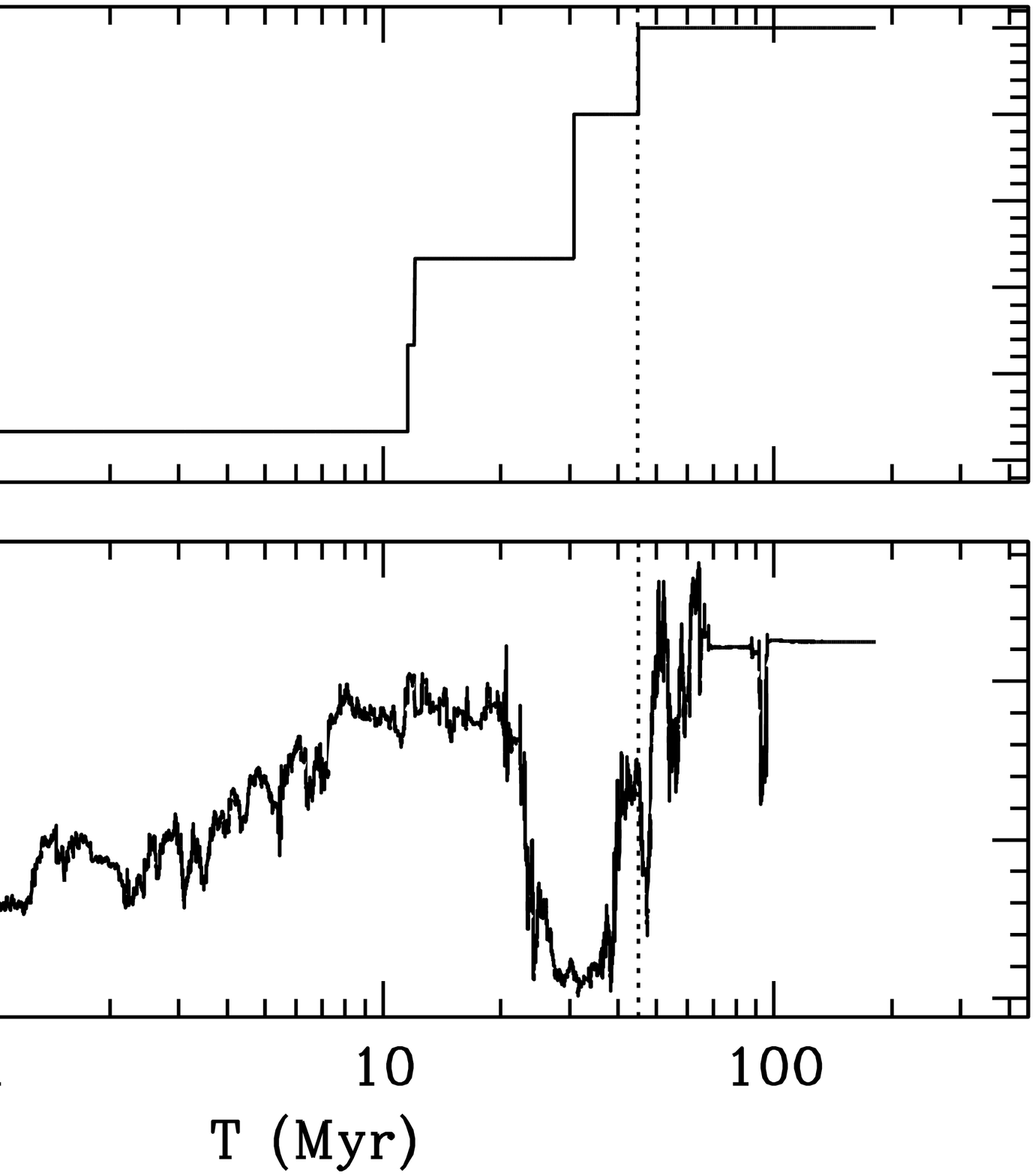}} \\
    \subfloat[]{%
     \includegraphics[width=.3\textwidth]{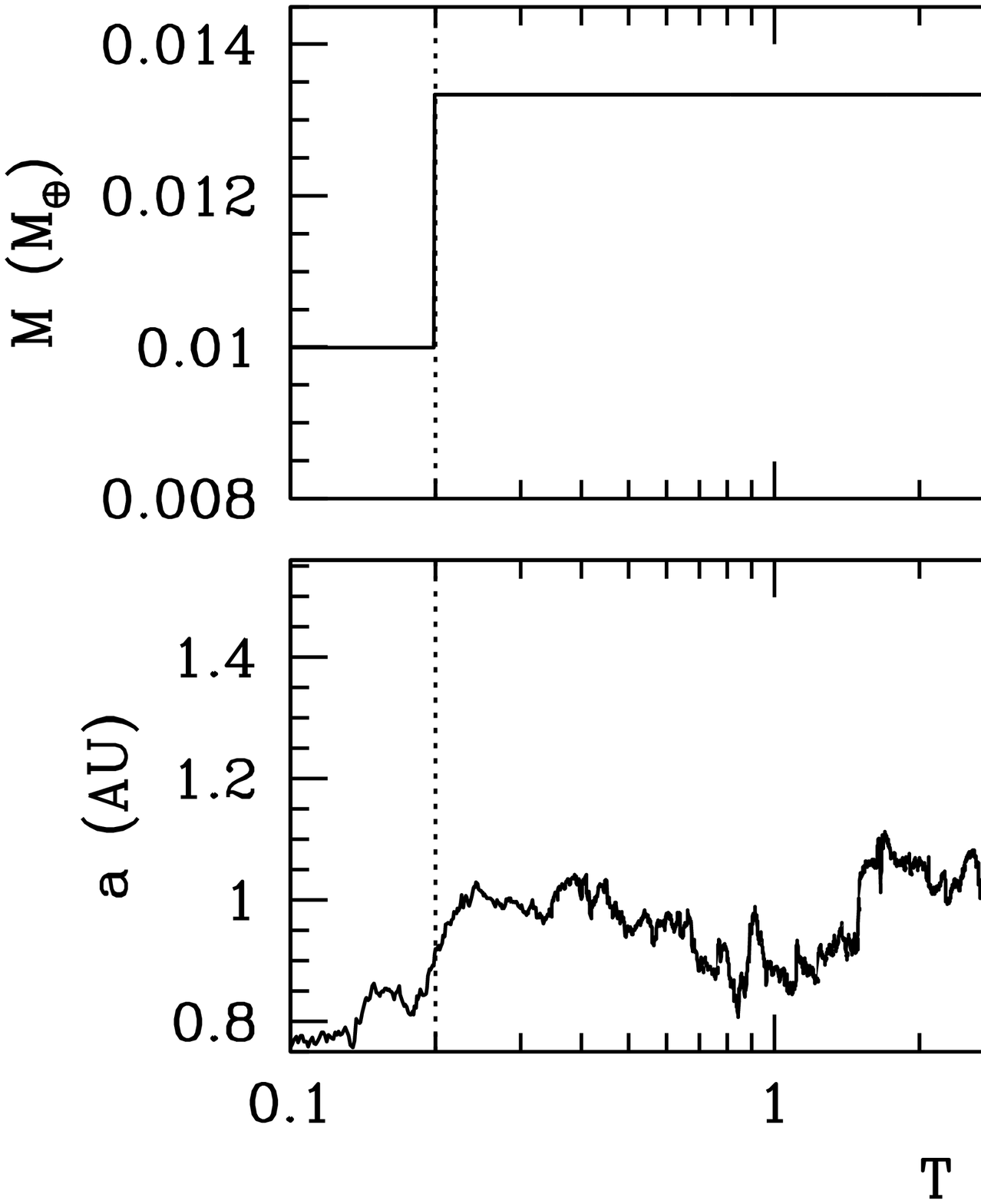}}\hfill 
    \subfloat[]{%
     \includegraphics[width=.3\textwidth]{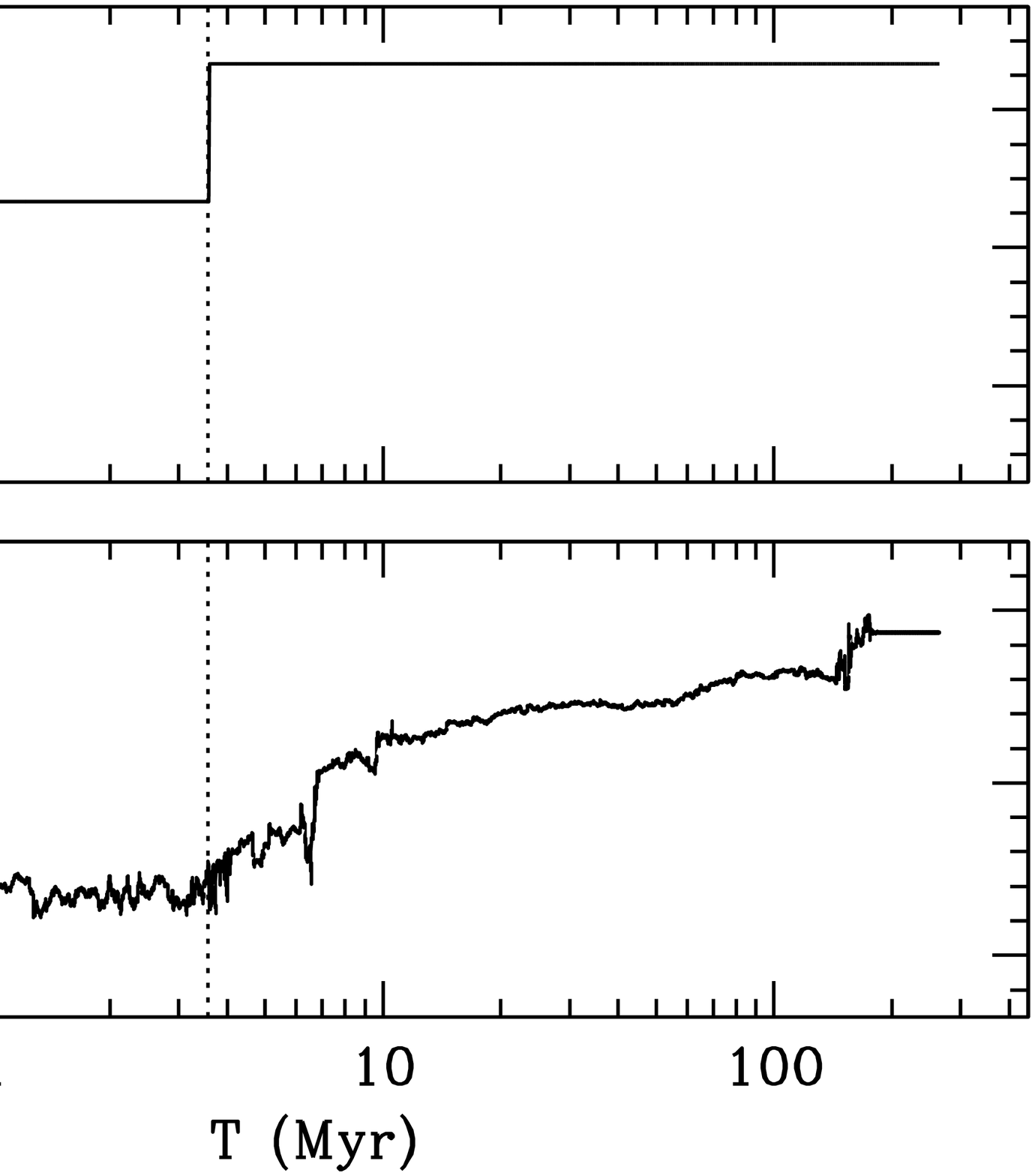}} \\
    \subfloat[]{%
    \includegraphics[width=.3\textwidth]{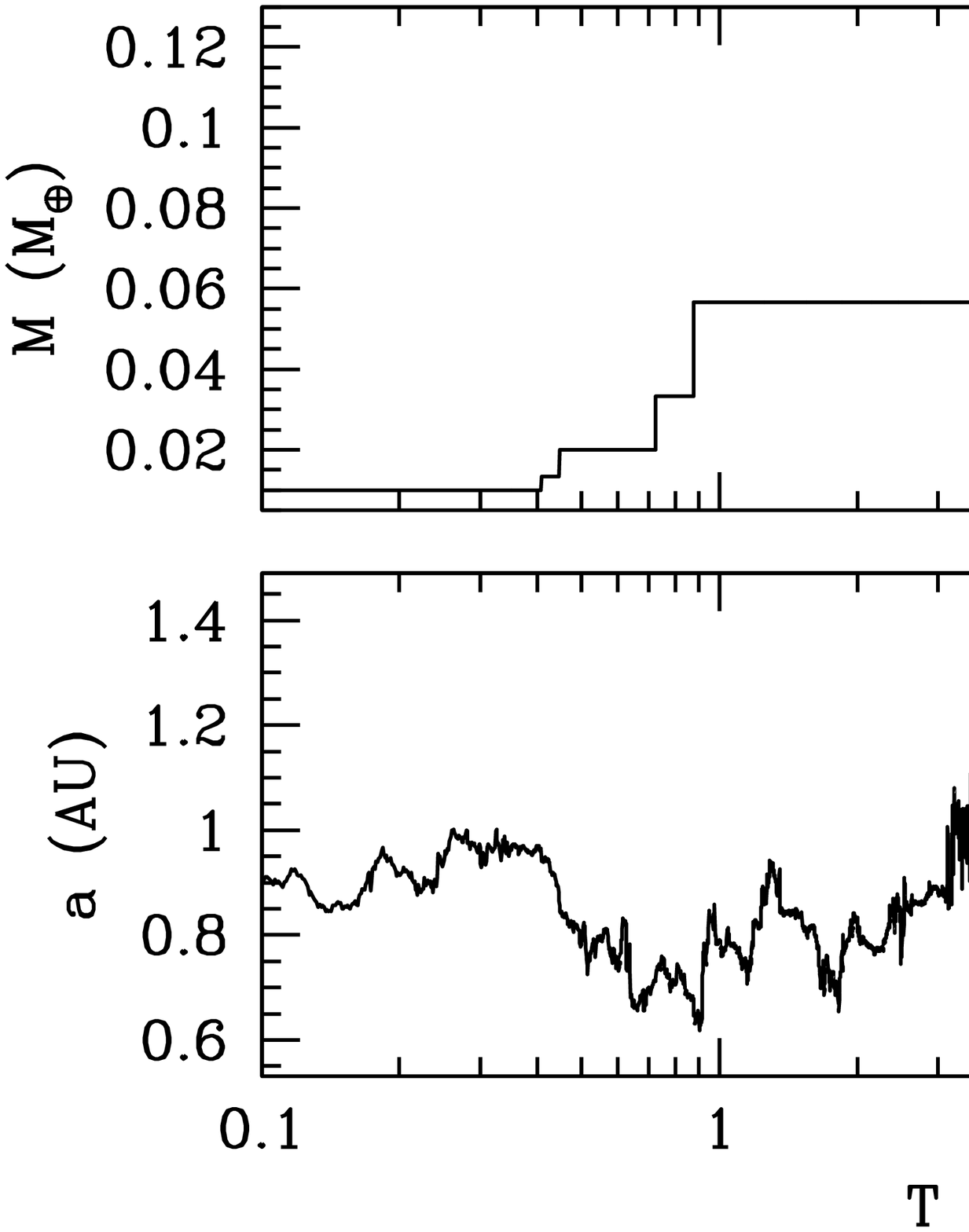}}\hfill
    \subfloat[]{%
    \includegraphics[width=.3\textwidth]{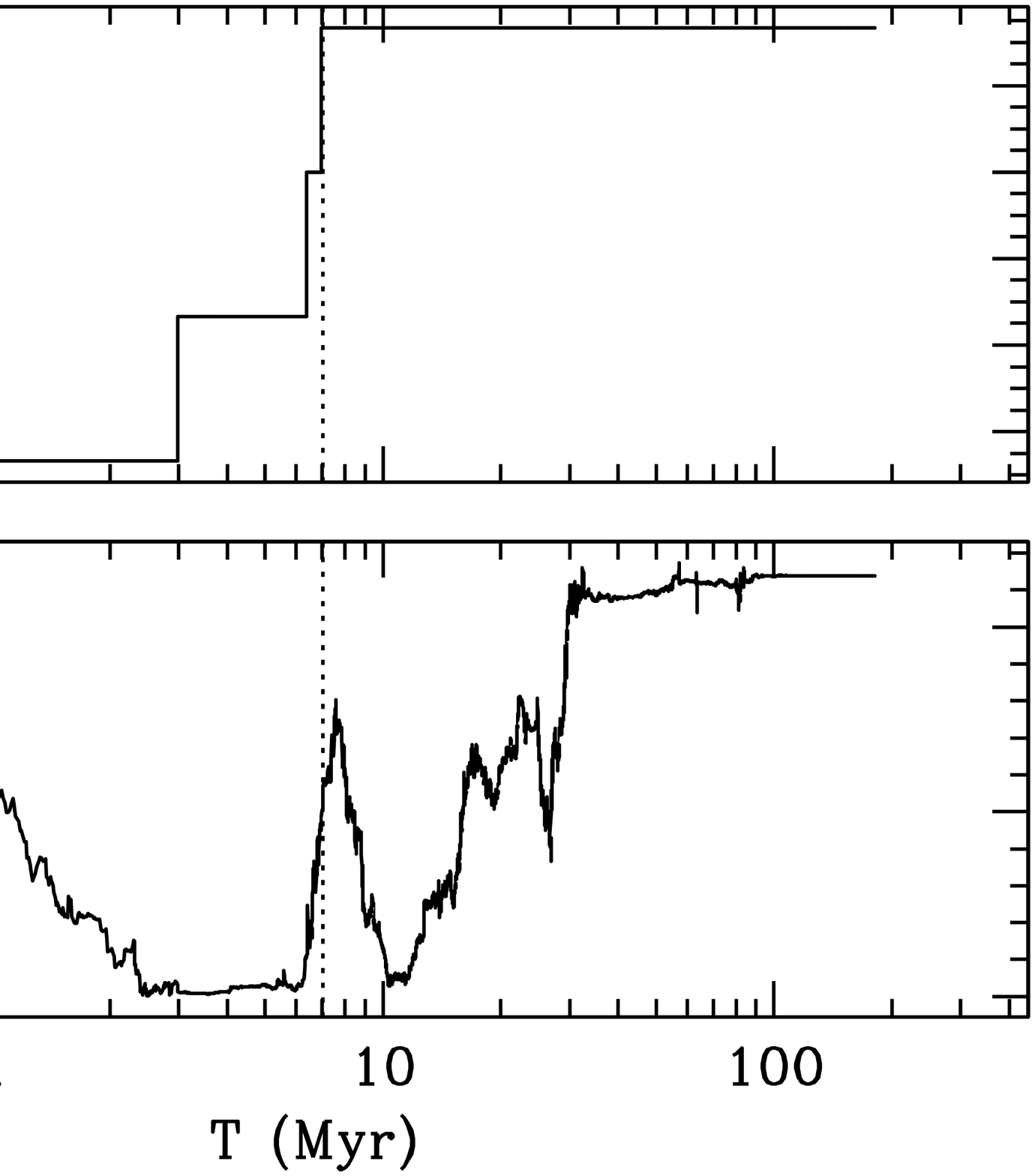}} \\
  \caption{This figure shows the assembly histories of six representative cases of Martian
analogues. In each case the upper panel shows the mass growth of the final TMA, and the vertical
dotted line shows the age at which it had the final impact with another body in the simulation. 
In each case the bottom panel shows the evolution of the semi-major axis on the same timescale.
Pabel a) shows the evolution of the final TMA designated Case~A (\#42). We see that the semi-major 
axis continues to evolve well after the last giant impact.
Panel b) shows the assembly history of Simulation \#30 (Case~B). Once again, the evolution to large
semi-major axis takes more than 100~Myr.
Panel c) shows  case~C (\#5).
 This body assembles quickly as it is only a little larger than
the original embryo size but then continues to diffuse in semi-major axis until
finally becoming dynamically decoupled at 65~Myr.
Panel d) shows case~D (\#10),
wherein the mass assembly finishes rapidly, but the semi-major axis evolution, shown
in the lower panel, continues for more than 100~Myr, eventually leaving the body with
a semi-major axis of 1.94~AU.
Case~E (\#3), shown in Panel e,
lasts for considerably longer than the cases shown in Panels a)--d).
The semi-major axis evolution (lower panel) also shows an outward diffusion, albeit somewhat more muted than
the other cases. The biggest qualitative difference here is that the evolution is terminated by a giant
impact, with no further evolution by gravitational scattering.
Pabel f) shows Case~F (\#HR6), which is one of the most distant, low mass of our true TMA. It is also one
of the high resolution runs, and so the qualitative similarity of the results demonstrates
our results are not sensitive to the numerical resolution.
\label{Histories}}
\end{figure*}

Figure~\ref{Histories} shows a selection of evolutionary histories for six of these TMA.
The first, denoted Case~A in Figure~\ref{MarsCases}, 
 is one of the closest to the observed Mars
in terms of mass and semi-major axis. The final mass is 
 $0.135 M_{\oplus}$, semi-major axis $a = 1.497$~AU. The eccentricity is a little
larger than the observed Mars (0.168) but the inclination of $4.5^{\circ}$ relative
to the original embryo orbital plane is  close to the observed inclination
relative to the Solar equator. The upper panel of Figure~\ref{Histories}a) shows the
mass assembly history of this body, with the bulk of the mass growth occurring within 10~Myr,
and the last major impact at 17.6~Myr. At this point the semi-major axis is 
 $\sim 1.4$~AU, but the TMA is still very much coupled to the rest
of the evolving planetary system -- crossing the orbits of several bodies and scattering
off of them. We see that the semi-major axis continues to evolve  for another 50~Myr, through
a long sequence of close gravitational scattering events which do not involve collisions
but do slowly cause the planetary orbital parameters to evolve.

Panel~b) of Figure~\ref{Histories} shows the history of case~B in Figure~\ref{MarsCases}. This
is less massive (final mass $0.06 M_{\oplus}$) and further out (semi major axis 1.62~AU)
than Case~A, but shows a similar history. In this case the mass growth occurs somewhat
later (last giant impact at 45~Myr) but the semi-major axis is still $<$1~AU at this
point. The outward migration is again driven by gravitational scattering after this, with
substantial changes in orbital parameters still occurring at ages $\sim 100$Myr.
Panel~c) shows the evolution of case~C in Figure~\ref{MarsCases}, which is
representative of the smallest of our likely TMA. This represents an even
more extreme case of the trend noted above, with planetary assembly complete by 0.2~Myr,
followed by 65~Myr of diffusion in semi-major axis until the final location is reached.

What about those objects located around the edge of our region of likely TMA?
The objects at the lowest masses comprise bodies that never underwent a planetary collision
and so are still of the same mass as the initial embryos. Their evolution is qualitatively
similar to that of case~C in Figure~\ref{Histories}, and so we do not show a figure for this.
 Panel~d) in Figure~\ref{Histories} shows the evolution
of Case~D, which has a similar mass to Case~B, but is located somewhat further out (1.94~AU).
The result is similar to other cases, with assembly complete after 3.6~Myr, followed by
an extended period of outward diffusion that lasts for $> 100$Myr. We also show panel~e),
representing Case~E in Figure~\ref{MarsCases}, which illustrates the inner boundary for the 
behaviour described here. This evolution is qualitatively distinct from those above, in that the
evolution in semi-major axis is truncated by a giant impact at 105~Myr, with no subsequent scattering
off any more bodies. Prior to this event, the evolution was qualitatively similar, and the difference
in outcome is a consequence of the fact that the last body keeping the outermost planet coupled to the
original annulus is removed by collision with that body, rather than one of the inner planets. Such evolution
can also characterise a handful of the TMA as well, primarily those towards the upper left of the box
in Figure~\ref{MarsCases}. Finally, case~F shows the history of an outcome at the lower right of our
allowed range, with a similar history of rapid assembly and long outwards diffusion. This simulation
was (\#HR6) was performed at higher resolution and still reproduces the qualitative behaviour.

\subsection{Mechanism of Decoupling}

The different histories shown in the previous section indicate that there is a qualitative difference in
the dynamical evolution of the Mars-like bodies and those that remain closer to the original annulus region -- namely
whether they experience substantial further evolution due to gravitational scattering after their last
giant impact. The outward evolution of the semi-major axis is driven by the transfer of binding energy to
the orbits of other bodies and will continue as long as objects of substantial mass cross the orbit.
For the final orbit to be stable in the long term, the orbit of the TMA must be decoupled from the annulus
from which it originally migrated. 
If the last interaction of the TMA is not a direct impact with another body, then there must
be another process to remove objects from crossing orbits. Here we wish to describe the process that 
is responsible for the dynamical decoupling of the TMA in each case.

\begin{figure}
\includegraphics[width=84mm]{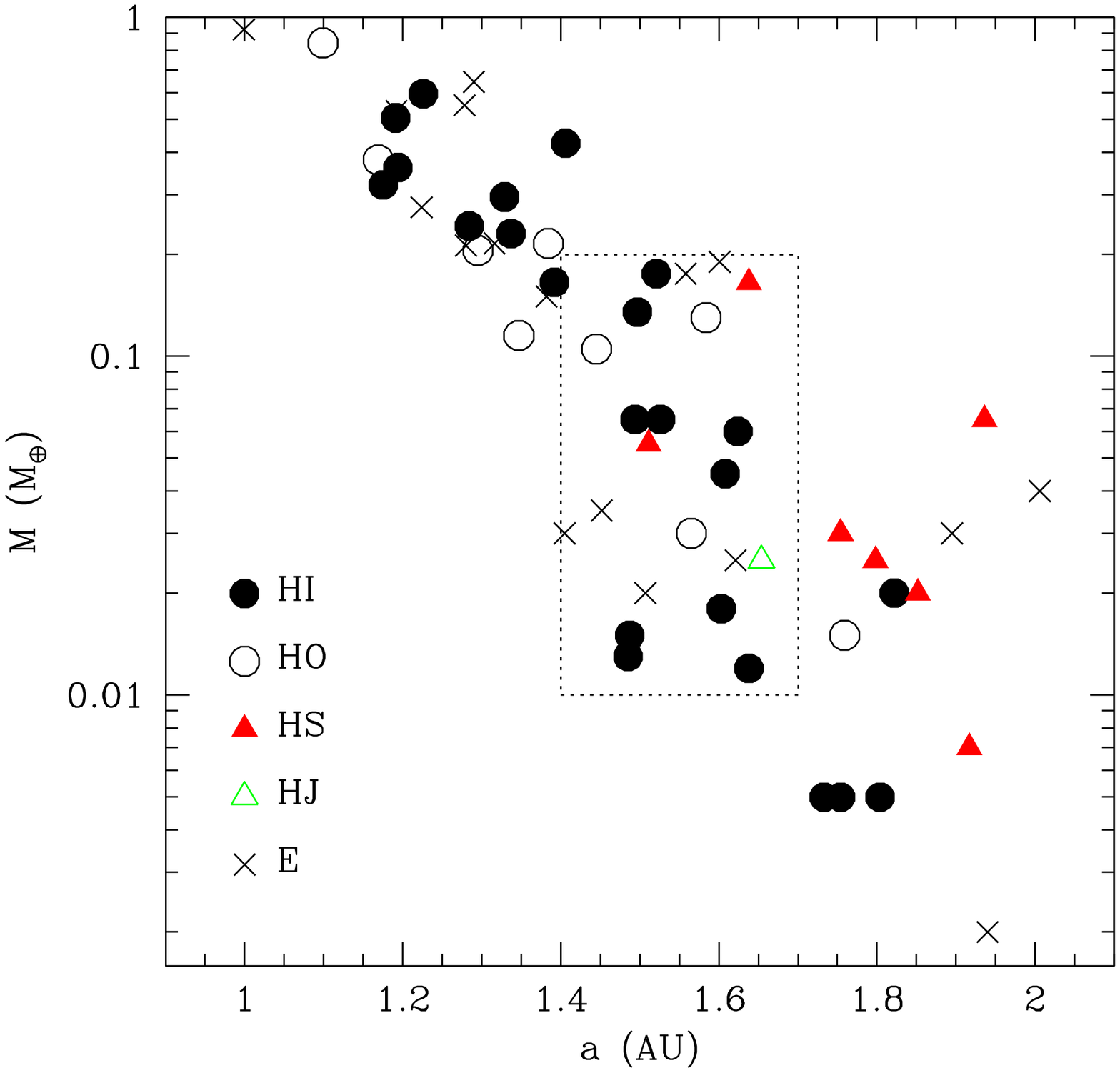}
\caption[How.ps]{The points show the semi-major axis and mass of each of the outermost terrestrial planets
in our simulations. The points are colour-coded according to the manner in which they finally became decoupled
from the larger inner planets. Black filled points (HI) indicate that the final removal was by a collision with one
of the larger inner planets. Crosses (E) indicate that the last body removed was ejected from the system.
Open points indicate systems in which the outermost terrestrial planet experienced the last collision (HO).
Red filled triangles indicate that the last planet removed hit the Sun (HS). The sole green open triangle shows the case
where the last collision was actually with  Jupiter. The dotted box indicate the regime we consider to
be likely TMA.
\label{How}}
\end{figure}

To that end, we present, in Table~\ref{Outcomes} and Figure~\ref{How}, the parameters relevant to the history of the outermost
surviving terrestrial planet in each of our simulations. We describe the final mass, semi-major axis,
eccentricity and inclination (relative to the original orbital plane) for each object, along with the time of 
the last impact experienced by the TMA and the manner in which the last body keeping the system dynamically coupled was removed.
If this object was removed by indeed striking the outermost of the terrestrial planets, we denote this
as ``HO'' (hit outermost planet).
If the last body was removed by striking one of the other terrestrial planets (i.e. not the outermost)
we denote this outcome as ``HI'' (hit inner planet). If the body was removed by impacting the Sun, we
denote this as ``HS''. If it struck the Jupiter it is designated as ``HJ'', and if it was ejected from the system entirely, this is denoted as ``E''.
We also indicate whether we consider this body a TMA. We denote by `L' (Likely), those that fall within
the dashed rectangle in Figure~\ref{MarsCases}, `P'(Possible), for those that lie outside this box but still have
$a>1.4$~AU and $M < 0.2 M_{\oplus}$, and 'No' for those that are either too massive or too close.

\begin{table*}
\centering
\begin{minipage}{140mm}
\caption{ Factors that characterise the encounter history. The
impact parameter b indicates the closest approach distance of any
object from the simulation (all) or by one of the planetary
bodies that survive to the end of the integration (big). $f_{disk}$ is the
fraction of the last impact-generated disk that remains bound at the 
end of the encounter history. $f_{cap}$ is the fraction of unbound
test particles captured in the last three body encounter experienced by
the TMA. $f_{cap} (q>1)$ is the fraction that survive their first
periareion passage.
 \label{EndPoints}}
\begin{tabular}{@{}lcccccccc@{}}
\hline
 $\#$ &
Mass & Semi-Major Axis &  Last Impact &  b (all) & b (big) & $f_{disk}$ & $f_{cap}$ & $f_{cap}$ (q$>$1)\\
    &($M_{\oplus}$) & (AU) &  (Myr) & ($R_{\male}$) & ($R_{\male}$) & & $10^{-4}$ & $10^{-4}$\\
\hline
2 & 0.005 & 1.484 &  $\cdots$ &  1.1 & 6.2 & $\cdots$ & 0.26 & 0\\
5 & 0.015 & 1.487 &  0.2 &  2.3 & 5.3 & 0 & 0.74 & 0\\
7 & 0.030 & 3.043 &  23.6 &  3.8 & 7.4 & 0 & 2.46 & 0.12\\
9 & 0.005 & 1.804 &  $\cdots$ &  1.1 & 6.7 & $\cdots$ & 0.25 & 0 \\
10 & 0.065 & 1.936 & 3.6 &  1.9 & 6.2 & 0 & 1.08 & 0.55 \\
12 & 0.005 & 1.754 &  $\cdots$ & 1.4 & 6.7 & $\cdots$ & 0.26 & 0\\
13 & 0.025 & 1.654 &  6.0 &  1.4 & 17.4& 0.13 & 1.59 & 0\\
15 & 0.045 & 1.608 &  7.4 & 1.7 & 6.1 & 0 & 0.7 & 0.2 \\
18 & 0.030 & 1.754 &  23.1 & 2.3 & 7.9 & 0 & 1.4 & 0.04\\
20 & 0.015 & 1.713 &  1.3 & 1.8 & 5.5 & 0 & 0.78 & 0\\
21 & 0.005 & 1.733 &  $\cdots$ &  1.2 & 3.7 & $\cdots$ & 0.27 & 0\\
23 & 0.020 & 1.852 &  3.2 &  1.7 & 3.1 & 0 & 1.58 & 0 \\
24 & 0.020 & 1.822 &  0.8 &  1.7 & 10.1 & 0 & 1.14 & 0\\
25 & 0.030 & 1.405 &  3.4 &  1.7 & 3.4 & 0 & 0.57 & 0 \\
26 & 0.065 & 1.526 &  62.6 &  7.4 & $\cdots$ & 0.67 & 1.85 & 1.25\\
27 & 0.165 & 1.638 &  30.7 &  2.9 & 10.5 & 0 & 0.06 & 0.04 \\
28 & 0.175 & 1.558 &  15.6 &  2.3 & 85.3 & 0.8 & 0.03 & 0.03\\
29 & 0.020 & 1.507 &  18.3 &  2.2 & 11.3 & 0.004 & 1.08 & 0\\
30 & 0.060 & 1.624 &  45.0 &  3.5 & 6.5 & 0 & 1.23 & 0.52 \\
33 & 0.065 & 1.494 &  45.5 &  2.6 & $\cdots$ & 0.46 & 0.10 & 0.04 \\ 
34 & 0.030 & 1.895 & 2.0 & 1.7 & 3.9 & 0.0 & 0.1 & 0.04\\
36 & 0.175 & 1.521 & 161.0 & 7.0 & $\cdots$ & 0.67 & 0.0 & 0.0\\
37 & 0.130 & 1.584 & 182.0 & $\cdots$ & $\cdots$ & 1.0 & $\cdots$ & $\cdots$ \\
39 & 0.015 & 1.719 &  788.3  & $\cdots$ & $\cdots$ & 1.0 & $\cdots$ & $\cdots$ \\
40 & 0.105 & 1.455 & 180.4 & $\cdots$ & $\cdots$ & 1.0 & $\cdots$ & $\cdots$ \\
42 & 0.135 & 1.497 & 17.6 & 1.9 & $\cdots$ & 0.17 & 0.1 & 0.07\\
44 & 0.040 & 2.006 & 28.5 & 2.6 & 7.9 & 0.0 & 2.47 & 0.75\\
45 & 0.025 & 1.798 & 1.4 & 1.6 & 3.2 & 0.0 & 1.02 & 0\\
46 & 0.030 & 1.565 & 780.4 & $\cdots$ & $\cdots$ & 1.0 & $\cdots$ & $\cdots$ \\
48 & 0.190 & 1.601 & 64.6 & 2.4 & $\cdots$ & 0.28 & 0.01 & 0.01\\
49 & 0.035 & 1.452 & 9.6 & 2.5 & 8.2 & 0 & 1.21 & 0.08\\
HR1 & 0.025 & 1.621 & 27.3 & 1.3 & 21.4 & 0.02 & 0.28 & 0.01 \\
HR2 & 0.002 & 1.940 & $\cdots$  & 1.2 & 6.9  & $\cdots$ & 0.01 & 0\\
HR5 & 0.018 & 1.603 & 14.3 & 1.9 & 3.2 & 0 & 1.16 & 0\\
HR6 & 0.012 & 1.638 & 6.9 & 1.3 & 3.8 & 0 & 0.75 & 0\\
HR8 & 0.013 & 1.485 & 26.5 & 2.9 & $\cdots$ & 0.65 & 0.73 & 0\\
HR9 & 0.055 & 1.511 & 14.3 & 2.2 & 16.0 & 0.14 & 0.91 & 0.33 \\
HR10 & 0.007 & 1.917 & 1.6 & 1.1 & 8.8 & 0.39 & 0.35 & 0 \\
\hline
\end{tabular}
\end{minipage}
\end{table*}

We see that the most common pathway to final decoupling is by
 by having the last interloping 
body impact one of the inner planets. This occurs in 25/60 of the simulations, and
in 10/22 of the simulations with a likely TMA. The next most common pathway is ejection
from the system entirely (17/60 and 6/22 respectively). In 9/60 cases, the outermost
terrestrial planet is the one that experiences the last collision, and three of these
cases actually fall within our preferred parameter space. Some of the planets are also
removed by collision with the Sun (8/60 and 2/22) and one is even removed by collision 
with Jupiter. This implies that most of the bodies that end up with $a>1.4$AU (31/35)
diffuse outwards long after their mass growth ends, due to  scattering off one or more intermediate bodies that cross their orbits and
that of the original annulus, where the larger assembled bodies continue to reside.
The larger cross-sections of these interior objects mean that they dominate the collision
probability and are therefore responsible for most of the final decouplings. This
 is also the reason why the mass accretion is much lower for the TMA
as they start to diffuse outwards -- where they no longer encounter as large a 
cross-section for collision as before.

\subsection{Impact Parameters}

The relevance of this outwards diffusion for the formation of Phobos and Deimos is twofold. If the
moons were formed as the
 result of a giant impact, this episode frequently occurs early on (at ages $< 10$ Myr) in the
planetary assembly phase. The fact that the planet continues to inhabit a dynamically
active environment for $\sim 100$~Myr then implies that the products of that collision 
will be subject to a large number of perturbations from planetary embryo mass (and larger)
bodies passing within the Hill sphere of the migrating TMA. This will provide a
substantial perturbation to the disk and can potentially destabilise it. However, it is not
clear that the disk will be completely destroyed. The encounters are significantly hyperbolic,
with relative velocities of several $km/s$, so that the time it takes the perturber to cross
a distance $\sim 10 R_{\male}$ is only $\sim 10^4$s, i.e. only $\sim 5\%$ of the orbital time
at this distance. This implies that the perturbations should be primarily in the form of direct
scatterings, with limited secular effects. As such, particles on the opposite side of the TMA
from the point of closest approach will be relatively weakly perturbed.
 On the other hand, there are potentially many such
perturbations. In \S~\ref{DiskKick} we will examine the susceptibility of moon forming disks
to the perturbation histories from the above simulations.

Alternatively,
if the moons are the consequence of capture, the series of late close passages with
other embryos offers an opportunity for three-body capture that does not exist in the
traditional scenario where Mars accretes more-or-less in situ and has to capture 
objects from the asteroid belt on a parabolic Keplerian orbit. This can potentially make the odds of capture
significantly higher as the varying gravitational potential provides a natural dissipation
mechanism, avoiding the need for invoking atmospheric or nebular gas drag.
In \S~\ref{Capture} we will examine the efficiency
of this capture process.

Table~\ref{EndPoints} shows the closest approach by another body experienced
by each TMA (P or T) after the last giant impact. This is expressed in
units of Mars observed radius ($R_{\male} = 3.39\times 10^5$~km). We see that nearly all experience the
passage of an embryo body within Deimos' orbit, and most experience one within the
semi-major axis of Phobos' orbit. Thus, the prospect for some level of interaction between the
nascent moon system and the perturbers is
almost guaranteed.
 We also record the closest passage with respect to one
of the larger bodies that ultimately survive to the end of the simulation. These have much
larger masses than the embryos and so contribute disproportionately to the integrated
level of disturbance experienced by any Moon-forming disks. Many TMA also experience an
encounter of this type on scales of Deimos orbit, although a few are fortunate enough to
avoid them.

\section{Consequences for Moon Formation}
\label{Consequences}

The evolutionary histories described in the previous section  provide the context for the
formation of any moon system surrounding a TMA. The dynamical perturbations experienced
via repeated close encounters will act to disturb and disrupt the disks of debris left behind
from a giant impact. If some fraction of the disk mass is able to survive, this could provide
a reason for why the Martian moons are so much lower mass than the Earth's.
On the other hand, if the perturbations are sufficient to remove the disk entirely, then we
will need to find another mechanism to create the moons. These orbital histories may actually
provide such a mechanism as well. In this event, the three-body interactions may provide an
alternative mechanism to supply the Moons via a capture process.

In this section we will examine the orbital histories of the above simulations to assess the
viability of these two pathways to Moon formation.

\subsection{Dispersal of Pre-existing Debris}
\label{DiskKick}

The conceptually simplest model for the formation of Martian moons is to start with the same
processes that are believed to be responsible for the formation of the Terrestrial Moon in
a giant impact (reviewed in Stevenson 1989; Canup 2004; Asphaug 2014). If we postulate that the last giant impact experienced
by each TMA produces a disk of debris that assembles to form a Moon (Hartman \& Davis 1975; Cameron \& Ward 1976;
Craddock 2011; Rosenblatt \& Charnoz 2012; Citron, Gendra \& Ida 2015; Canup \& Salmon 2016; Rosenblatt et al. 2016; Hyodo et al. 2017), then we can
examine the degree to which the continued perturbations resulting from the close passages
of other bodies will disturb such a disk. This leads to the hypothesis that the much smaller mass
of the Martian moons is due to the dynamical erosion of the original reservoir.
 A detailed calculation of moon formation would require a substantial treatment of not
only the gravitational interactions but also the chemical, hydrodynamic, radiative and collisional aspects
of impact products and their evolution, which is beyond the scope of this investigation. However,
in this section we wish to at least quantify the degree to which the disk is
likely to be affected by the gravitational perturbation environment.
This can then be used to  frame  a more detailed investigation.

\begin{figure}
\includegraphics[width=84mm]{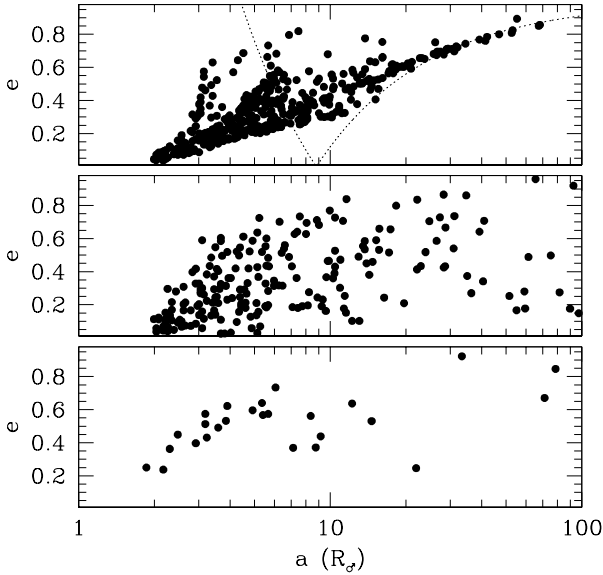}
\caption[Debris2.ps]{The upper panel shows the eccentricity--semi-major axis distribution of the model debris disk
in simulation \#5 (Case~C) 
after a single close passage of a $0.015 M_{\oplus}$ perturber. The dotted lines indicate orbits that have periareia
and apareia at 8.8$R_{\male}$, the impact parameter of the encounter.
 The middle panel shows the surviving population
after 10 such passages, as dictated by the simulation orbital history. The next encounter was with a much larger
body, and leaves a substantially denuded population, shown in the bottom panel. These remaining objects are then
also lost in subsequent encounters.
\label{Debris2}}
\end{figure}

To that end, we performed the following idealised calculation. We again
use the {\tt Mercury} code, but now will use the TMA in each case as the central mass.
For simplicity, we will assume all central objects in this section have the
same physical size as Mars. We extract
from the data on close encounters (for that portion of the history that comes
after the last giant impact) in each TMA history the relative position and velocities, as well as the perturber
mass, for each passage of a massive body within the TMA Hill sphere.\footnote{ In cases where the recorded values start the perturber too close
to the central body we perform a brief backwards integration to an earlier starting point, in order to properly resolve the entire passage through the Hill sphere.}
We then integrate a population of 500 test particles, initially spread uniformly in semi-major axis from 2--10$R_{\male}$ and on circular
orbits in the planetary equatorial plane. We also use the Martian value of  $J_2=1.97 \times 10^{-3}$ to represent the flattening of
the central body. This will allow us to
properly capture the effects of precession. Our goal here is not to provide an accurate model of a debris disk
(initial debris disks are likely much more compact and collisional) but rather to illustrate the susceptibility of 
debris to disruption by passing bodies. The assumption of a primarily gravitational evolution is justified
by prior analyses (Thompson \& Stevenson 1988; Machida \& Abe 2004; Charnoz \& Michaud 2015), which suggest that the timescales for condensation, cooling and solidification of a
protolunar disk are much shorter than the intervals between the encounters considered here.

The TMA with the longest exposure to the dynamical elements is found in the simulation
we designated as Case~C above. In this case, the last major impact occurred at 0.2~Myr, so that the
resulting debris disk is exposed to the perturbations by passing bodies for 46~Myr (the time until the
last close passage by any body within 10 $R_{\male}$ in this case). Figure~\ref{Debris2} shows how the
model disk described above evolves under these conditions. The upper panel shows the state of the
disk after the first passage, which is by a $0.015 M_{\oplus}$ body with an impact parameter at 8.8$R_{\male}$.
The disk is clearly disturbed, but not disrupted by any means. The dotted lines in the upper panel show the
orbits that cross a circular orbit at the impact parameter of the encounter. Points between these curves are
consistent with an impulsive scattering event. The slope of the distribution interior to this can be understood
as the impulsive response to the high velocity passage of a perturber. Integrating the effect of a perturber
passing in a straight line yields $\delta e \propto a^{3/2}/b^2 V_0$, where the constant of proportionality
depends on the position of the particle relative to the point of closest approach.
 The approximation of large $V_{\infty}$ is only weakly satisfied on the scale of Phobos' orbit, 
but is sufficient to motivate the observed trend of eccentricity with semi-major axis.
 Individual encounters by small bodies
such as this (only a factor of a few larger than the initial embryo mass) are the most common kind of interaction and are
not strong enough to completely disrupt the disk in a single encounter. However, the
accumulated  result of many scatterings
is enough to substantially heat the disk in a dynamical sense.

The middle panel shows the system after
10~encounters with such bodies, all of mass between $0.005$--$0.015 M_{\oplus}$. At this point it has been
whittled down to 223 bound objects (so 45\% of the original population survive), although the surviving
population is now substantially excited dynamically. At this point in the simulation (only 0.2~Myr after
the impact) the encounter is with a much larger object (0.12$M_{\oplus}$), which unbinds the bulk of
the remaining bodies. The bound population after this encounter is shown in the bottom panel.
Some of these remaining fragments survive a handful more encounters, until a passage by an even bigger
body ($0.56 M_{\oplus}$) at 0.8~Myr removes all of the survivors. This is therefore an example of a TMA history
that retains no remnant whatsoever of its last impact-generated population of satellites. 

\begin{figure}
\includegraphics[width=84mm]{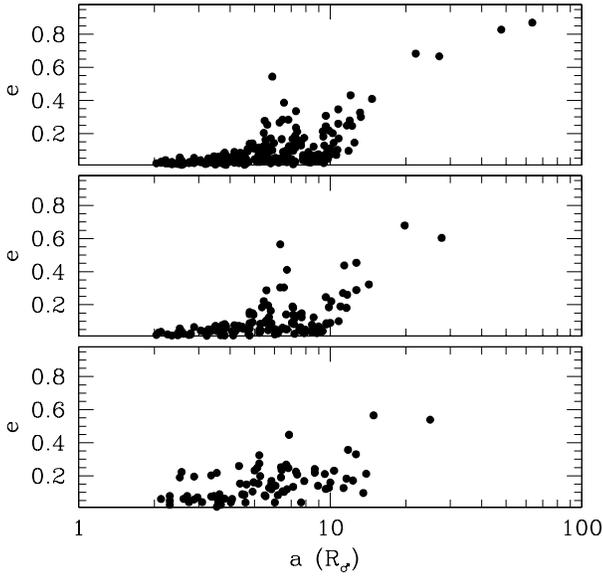}
\caption[Debris4.ps]{The upper panel shows eccentricity versus semi-major axis after the first significant
encounter at 28.6~Myr, for the history of Simulation~\#42. At this point the bound mass is 48\% of the original. The middle panel shows the disk
at 38.7~Myr, after the next significant scattering event, when 29\% of the original mass is still bound. The
bottom panel shows the final state of the disk, after dynamical decoupling, in which 17\% of the mass is
still bound.
\label{Debris4}}
\end{figure}

A less extreme kind of history is shown by the TMA in Case~A. In this case the final
giant impact occurs  later, at 17.6~Myr, and occurs when the semi-major axis is already 1.4~AU,
and so the dynamical perturbations of the impact-generated disk
 begin with the TMA already removed from the region of majority mass concentration. The result of this is that the 
dynamical evolution is less violent. Although there are multiple close passages by embryo-size
bodies, there are no passages close to the largest bodies remaining in the annulus. A series of snapshots
of the resulting evolution is shown in Figure~\ref{Debris4}. The evolution is dominated by a handful
of particularly close encounters -- a 0.015$M_{\oplus}$ body passing within 5.7$R_{\male}$ at 26.5~Myr,
a 0.005$M_{\oplus}$ body passing within 6$R_{\male}$ at 38.7~Myr, and the passage of a 0.01$M_{\oplus}$
body within 2.6$R_{\male}$ at 46.2~Myr. The final surviving population, shown in the bottom panel of Figure~\ref{Debris4}
retains 17\% of its original members, in a somewhat dynamically excited state.

Of the 34 systems summarised in Table~\ref{EndPoints}, 16 experience so many dynamical perturbations
after their last giant impact that their disks are completely removed (to the level of the resolution
used here). A further 9 retain a reduced and dynamically excited disk at the end of the simulation,
with a wide range of retention fraction which depends on the particular environment and history. In four
cases, the giant impact was the last significant dynamical encounter the TMA experienced, and in five
cases the TMA was a surviving planetary embryo, in which case it had no giant impacts whatsover. 
Thus, we see that there is a wide range of potential outcomes spanning the full gamut from no giant
impacts at all, to scenarios that mimic the origin of the Earth's moon directly. Of most interest,
however, is that 74\% of the TMA origin scenarios result in at least some level of disturbance to the
putative protolunar disk.

\begin{figure}
\includegraphics[width=84mm]{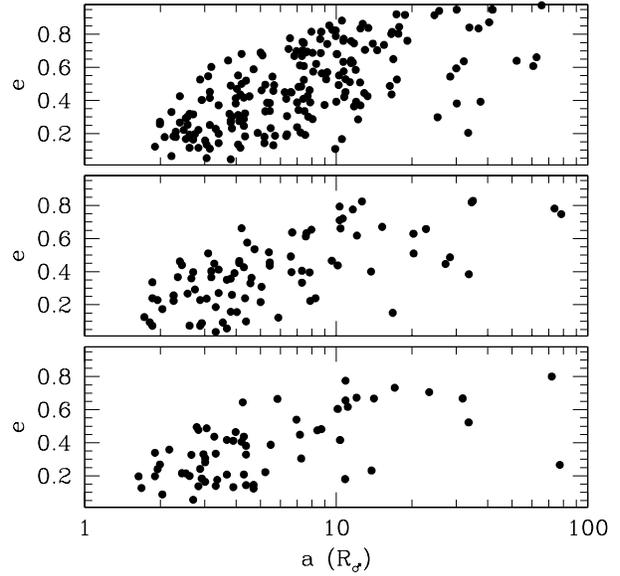}
\caption[Debris3.ps]{The upper panel shows inclinations versus semi-major axis after 10.4~Myr,
after the post-giant impact disk has experienced 10 close passages of a massive body within
10~$R_{\male}$. At this point 204 bodies remain bound to the TMA. The middle panel
shows the state of the disk after 15.8~Myr, at which points 91 bodies remain. As the system
evolves towards dynamical stability, the encounters become less frequent, and the lower panel
shows the disk after the last close encounter, at 103.9~Myr, at which point 67 bodies remain.
This represents the encounter history from Simulation~\#13
\label{Debris3}}
\end{figure}

 Figure~\ref{Debris4} showed an example of a disk that survived an extended period of disturbance, in
which the erosion was dominated by a few significant events.
Another  case (simulation \# 13) is shown in Figure~\ref{Debris3}. In this case, the $0.025 M_{\oplus}$
TMA experiences its last collision at 6~Myr, and so still has a substantial number of close 
passages to navigate. However, none of these are with a body substantially more massive and so the
decline of the population is characterised by a steady erosion, rather than a few catastrophic events.
Figure~\ref{Debris3} shows the evolution of the disk over the course of 39 successive passages inside
10~$R_{\male}$. At the end, 67/500 particles remain, again in a dynamically hot state. 

It is histories such as this one that offer the most tantalizing prospect for a coherent origin
scenario for the Martian moons. It allows for the survival of a disk much less massive than what
one would naturally expect from a giant impact and yet still retaining some vestigial disk-like
properties that might help to explain the equatorial orbits of the moons. We will investigate
the further evolution of such surviving disks in \S~\ref{Aftermath}.

\subsection{Capture of Planetesimals by Three Body Interactions}
\label{Capture}

One question that hangs over the calculations in Figure~\ref{DiskKick} is whether a moon forming
disk can even form in collisions with a body as small as Mars. It is difficult to capture material
into a stable circumplanetary orbit by purely ballistic means -- most scenarios for formation of
the Moon rely on pressure gradients in vapour-liquid mixtures to slow material on otherwise hyperbolic
trajectories. Vaporization of rock requires velocities $\sim 7$km/s (Ahrens \& O'Keefe 1972), which is larger than
the
  escape velocity from the surface of Mars
($\sim 5$km/s). Nevertheless, modern simulations with the best available equations of state do produce
moon-forming disks (e.g. Hyodo et al. 2017), but the physics of such simulations are necessarily limited.
Furthermore, even if disks do form, in many cases they can be completely removed by dynamical perturbations,
as we have seen in the prior section.
Given these uncertainties, it is of interest to examine the pathways to capture for the Martian moons.

The gravitational perturbations from the close passage of another massive body can cause low mass bodies (test particles) passing by the planet
on hyperbolic orbits to become bound to the planet. This is simply the inverse of the process that  unbinds bodies
 in the previous subsection. Of course, this can occur only in a very limited range of parameter space, and so we do not
anticipate being able to reproduce such captures spontaneously from a full N-body simulation. Instead, we
recreate the passage of an individual encounter taken from our N-body simulations and integrate a large number
of test particle orbits on a range of initially unbound trajectories. Once the perturber has passed, we 
examine the parameters of the bound objects.

\begin{figure}
\includegraphics[width=84mm]{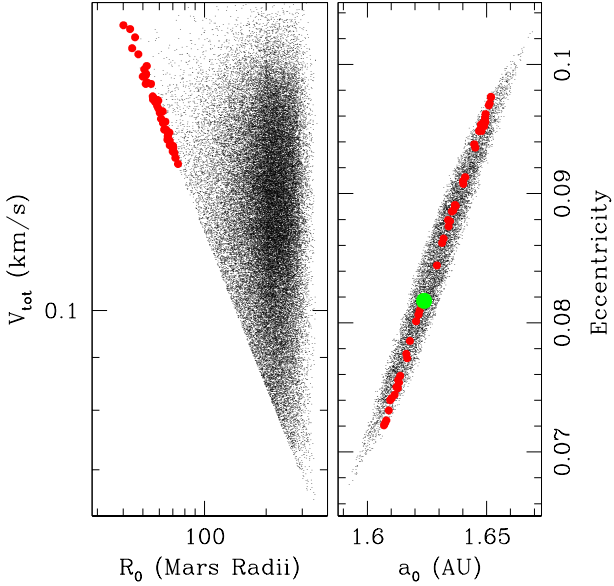}
\caption[Sample.ps]{The left hand panel shows initial distances and velocities of each
 test particle relative to the TMA, prior to the passage of the perturbing body, in
the case of the last encounter in simulation~\#30 (Case B).
Red points indicate those that became bound after integrating the passage of the perturber,
with a semi-major axis that lies within the TMA Hill sphere.
The blank space to the lower left indicates orbits that would have been bound at the start
of the integration. The right hand panel shows the heliocentric orbital elements corresponding
to the points in the left hand panel. Once again, the red points indicate those that became bound.
The green point indicates the orbital properties of the TMA at the time of the encounter.
\label{Sample}}
\end{figure}

Let us consider the capture process during the last encounter experienced by the TMA in each simulation. As an example, let us consider the case of
Case~B.
 We describe the motion of the perturber in each encounter, relative to the TMA, as a hyperbolic Keplerian orbit with the initial orbital position
and relative velocity derived from the simulations. For each orbit we  numerically integrate, with a Runge-Kutta
integrator, $10^6$ test particle orbits in the frame centered on the TMA and subject to the acceleration of the perturbing body. 
Integration is carried
out, with adaptive stepsize control, for the duration of the perturber passage through the Hill sphere and
the Keplerian orbital parameters are re-evaluated at the end of the integration.
 At these low
masses and high encounter velocities, only particles on very similar trajectories to the primary 
are likely to be captured. Thus, we choose initial starting positions for the test particles as randomly oriented
 relative to the TMA,
with a distance randomly chosen between zero and  the  Hill sphere radius,
 and with a randomly directed velocity of magnitude, relative to the TMA, between zero and 0.25~km/s. We exclude
any combinations of initial conditions that would produce a bound orbit at the start. 
Figure~\ref{Sample}
shows the initial radii and total velocities relative to the planet, as well as their heliocentric orbital
parameters, for each test trajectory. The red points indicate those initial conditions that eventually lead to bound orbits. 
As expected, only particles on very similar trajectories to the TMA 
are likely to be captured. The right hand panel of Figure~\ref{Sample} shows what these relative positions
and velocities mean in terms of heliocentric orbital elements and the green point
illustrates the orbital elements of the TMA.

For this case, out of $10^6$ integrations, 123 particles were captured. We only count as captured those
which have a final semi-major axis that lies within the Hill sphere of the TMA. Long term stability likely
requires an even more stringent criterion, but we will see that the captured population is likely to be
highly collisional and will evolve rapidly.
 Of the captured objects, 67 were on trajectories so eccentric that their 
 periareia would lead them to strike the surface of the planet upon a first passage, while 52 would
survive at least the initial passage. Of these remaining 52, 15 had periareia $> 2 R_{\male}$. Thus, we
expect the encounters to generate both a shower of small body impacts, as well as a weakly bound cloud
of bodies on highly elliptical orbits, some of which may become tidally disrupted as they pass close to
the planet. We will characterise the capture rate as the fraction of captures that avoid initial
impact (so $f_{cap} = 5.2 \times 10^{-5}$ in this case). This is not the global capture rate of planetesimals from the
background population, because it samples only a small part of the position and velocity space, but
it provides a metric to characterise the capture probabilities for different TMA. We will address the
global rate of capture in \S~\ref{Demographics}. We list the $f_{cap}$ for each simulation in Table~\ref{EndPoints},
as calculated above for the last encounter in each simulation. In cases where the last encounter is a
giant impact, we put $f_{cap}=0$.

\begin{figure}
\includegraphics[width=84mm]{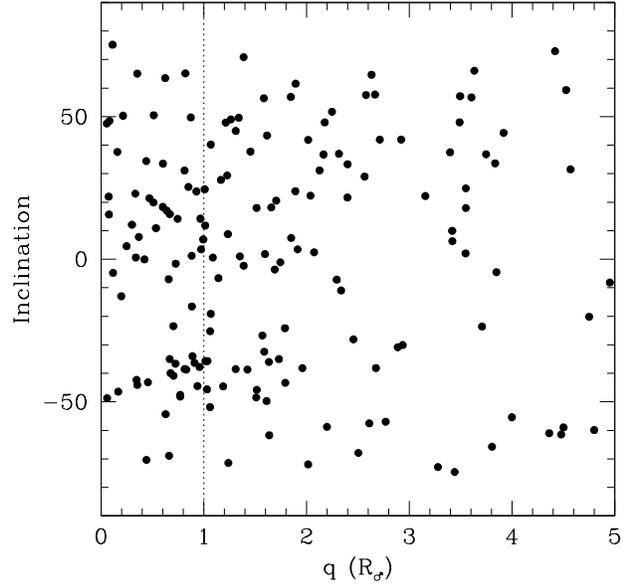}
\caption[Captured.ps]{The points indicate the Mars-centric orbital elements of the objects captured in
the last encounter experienced by the $0.065 M_{\oplus}$ TMA in simulation 26. We
show the periareion of each orbit versus its inclination relative to the orbital plane of the original
planetary system (any spin tilt incurred during accretion may change this). The vertical dotted line
represents the present Martian radius. We see that a significant fraction of orbits are likely to
impact the surface on a first pass, but that many will survive. This implies that the outcome of the
close encounter is a quasi-isotropic cloud of small bodies bound to the planet on eccentric orbits. 
\label{Captured}}
\end{figure}

The capture yield varies between cases. If we repeat the above calculation for the last encounters
in the cases with the
two most massive analogues (Case~A being one of them), we find a much lower yield of captures. In
Case~A, we find only 10 captures out of $10^6$ integrations, with 3 of those which should immediately
impact the planet. Other cases with more massive TMA also yield fewer captures (but with larger
fractions that survive the first pass).
 At the other end of the mass scale, the lowest mass TMA 
capture objects, but the orbits are so elliptical that all of the captured objects strike the planetary
surface on the first pass. Case~C is emblematic of this group, and we find 74 captures out of $10^6$
integrations, but no objects that survive a first pass. The most successful martian analogues are
those in the mass range 0.03--0.07$M_{\oplus}$. As an example of the captured population in one
of these cases, Figure~\ref{Captured} shows the 185 objects captured in $10^6$ integrations for the
case of Simulation~26, featuring a 0.065$M_{\oplus}$ analogue. We show the post capture periareia versus
orbital plane inclination (relative to the original plane of the protoplanetary disk). We see that 60
of these will impact the planet immediately, leaving 68\% of the captured bodies on longer-lived
orbits. Note also that the orbits are approximately isotropically distributed, with many of the captured
bodies orbiting in a retrograde sense.

\subsubsection{Demographics}
\label{Demographics}

What is a plausible number of captured objects? We have simulated only a small fraction of the velocity
space of potential encounters. To estimate what fraction of the total population of encounters this
represents, we assume that the velocity distribution of the small bodies is well matched to that
of the surviving embryos during the last stages of planetary clearing. In the absence of dynamical friction
amongst the small bodies, this should be a reasonable approximation. If there was dynamical friction, velocities
of small bodies would be damped and the yield potentially much higher.
 For each simulation, we examine
the relative velocities for the last 100 encounters and fit this to a three-dimensional Maxwellian
velocity distribution, parameterised by a dispersion $\sigma$. This provides a reasonable fit in many cases, 
although not all, but suffices for the following order of magnitude estimate. At the low
end of such a distribution, the fraction of the population contained below a velocity $V_0 << \sigma$ is
\begin{eqnarray}
f_v & = & P(V<V_0) = \sqrt{\frac{2}{\pi}} \frac{V_0^3}{2 \sigma^3} \nonumber \\
 & \sim& 1.7 \times 10^{-5} \left(\frac{V_0}{0.2 km/s}\right)^3
\left( \frac{\sigma}{5 km/s} \right)^{-3}.
\end{eqnarray}

Thus, for a given encounter, a fraction $\sim f_{cap} \times f_v$ of all nearby small bodies will be captured into
bound orbits. Given the values of $f_{cap}$ in Table~\ref{EndPoints}, this suggests a capture fraction $\sim 10^{-10}$ in the optimistic cases.
To estimate the size of the overall source population of small bodies, we note that observations of highly siderophile elements in
the crust of the terrestrial planets have been used to estimate the strength of the `late veneer', which is the amount
of chondritic material added by accretion after core formation of the differentiated bodies. Bottke et al. (2010) estimate
that Mars has accreted $\sim 2 \times 10^{24} g$ of material during this epoch.  In order to link this to our
background population of planetesimals, we note that directly accreted material must also pass through the Hill sphere,
and so we can use the accreted fraction $f_{lv}$ to normalise the total mass passing through the Hill sphere during
encounters. This factor is given by 
\begin{eqnarray}
f_{lv} & \sim & \left(\frac{R_{\male}}{R_{Hill}}\right)^2 \left( \frac{V_{esc}}{V}\right)^2 \nonumber \\
&\sim & 5 \times 10^{-6}
\left( \frac{a}{1.5 Au} \right)^{-2} \left(\frac{M_p}{0.1 M_{\oplus}} \right)^{1/3} \left(\frac{V}{5 km/s} \right)^{-2}
\end{eqnarray}
where $R_{Hill}$ is the Martian Hill sphere, $V_{esc}$ is the escape velocity from the Martian surface, and V is the
velocity of encounter (we have assumed gravitational focussing dominates). 

If we divide $2 \times 10^{24}$g by $f_{lv} = 5 \times 10^{-6}$, we find that a total of $67 M_{\oplus}$ has passed
through the Martian Hill sphere during the time of the late veneer deposition. This is, of course, an integrated
quantity, and represents multiple passages of the same bodies. To estimate the mass passing through the Hill
sphere {\em per encounter}, we note that our simulations show $\sim$~3000--8000 encounters within the Hill sphere
after the last giant impact (+ 10~Myr for core differentiation -e.g. Jacobsen 2005) for the different TMA considered here. Thus, each
encounter features $\sim 0.01 M_{\oplus}$. If we invoke the above accretion efficiency $\sim 10^{-10}$, we find 
approximately $\sim 10^{16}$g of material captured per encounter.

We can perform the same estimate for each simulation, using the appropriate $\sigma$, $f_{cap}$ and the number
of encounters post impact for each, but normalising, in each case, to the Bottke late veneer estimate. The
most optimistic scenario is simulation~26, where $\sigma \sim 4.5 km/s$, $f_{cap} \sim 1.25 \times 10^{-4}$ and there were 4604 Hill sphere
passages during the late veneer phase. This ultimately yields $1.6 \times 10^{18}$g per encounter. The next best
scenarios are \#~44 ($1.2 \times 10^{18}$g) and \#~15 ($2 \times 10^{17}$g). These all suggest that an individual encounter
will capture mass that is a fraction of Phobos or Deimos. It is, however, only a very approximate figure, as it depends
on the size of the background population and it's velocity distribution, both of which are quite uncertain. Furthermore,
 integrating over many encounters, the amount
of mass captured into this population can exceed this estimate by orders of magnitude. The principal question
is whether previous generations of material are unbound by each new encounter. The orbits shown in Figure~\ref{Captured}
are easy to disrupt, but collisional evolution of a swarm of small bodies may produce more strongly bound remnants
on a timescale short compared to the interval between encounters.

\section{Longer Term Evolution of Remnant Populations}
\label{Aftermath}

We have thus far demonstrated that the satellite population of Mars is likely
to have been heavily sculpted by the dynamical perturbations that Mars 
 experienced due to its outward migration from the birth annulus. There
is still a substantial amount of dynamical evolution between this epoch and the
presently observed population. A comprehensive exploration of this evolution is
clearly of great interest but also requires an extensive treatment of its own. 
Here we will restrict our attention to the more limited question of how the test
particle satellite populations identified in the previous section are likely to
be affected by collisions. This should help to frame more detailed calculations
in the future.

\subsection{Collision Evolution of Remnant Disk}
\label{Collision}
The simulations of \S~\ref{DiskKick} treated the disk as composed of test particles, as it was intended as
a simulation of susceptibility to disruption, rather than a genuine simulation of disk evolution.
Each test particle in such a disk could be properly considered its own realisation of a moon forming event.
Nevertheless, the separation of Phobos and Deimos suggests that material forming the Martian satellites was spread over
a range of several Martian radii. Thus, we will adopt the above disks as crude realisations of initial satellite
swarms, and examine how they might evolve as a collisional system.

 For a dynamically excited disk,
composed of bodies of similar size to Phobos (we take a mass $10^{19}$g and radius 10~km), the
collision cross-section is essentially geometric, and the relative drift of orbits of
incommensurate periods will bring them into contact on a timescale
$\sim 70 (R/R_{\male})^{3/2}$~days. The precession due to Mars $J_2$ will yield drifts of
similar magnitude (albeit with a steeper -- 7/2 -- power law). Therefore we expect these
disks to evolve rapidly due to collision.

 As a first examination of the consequences of this evolution, we have taken the remnant disks
in the four cases above where the disk was reduced to less than 20\% of its original mass, but
now evolved the remnant system treating the surviving bodies as massive bodies.
 In order to keep the consideration relative uniform,
we assert that each original disk was $\sim 0.01 \times 0.005 M_{\oplus}$ in mass (Marinova, Aharonson \& Asphaug 2011), although
impactor masses vary a little from one scenario to the next, and the mass is divided equally
amongst the original (before any perturbations were applied) 500 particles. This yields an individual particle mass of $\sim 6 \times 10^{20}$g.
This is still $\sim 60$ times larger than Phobos, but nevertheless representative of a disk that is of much
lower mass than expected from the undisturbed case.
 We assume densities of $2 g/cm^3$ for these bodies.

Once again, we integrate this population using {\tt Mercury}, setting the mass of the TMA
in each case for the central object, and using a Martian $J_2$. We also include the Sun as a distant perturber,
using the orbital parameters from the last encounter in each simulation. We run each system for $\sim 10^4$ years,
with a timestep of 0.01~days.
Table~\ref{GroundDown} shows the masses, semi-major axis, eccentricity and
equatorial inclination of each surviving body, as well as the inclination of
the orbital plane of the perturbing Sun.

\begin{table}
\centering
\begin{minipage}{140mm}
\caption{ Final states of disks and  swarms after collisional evolution.
 \label{GroundDown}}
\begin{tabular}{@{}lccccc@{}}
\hline
 $\#$ &
Mass & a &  e & I  &  I$_{\odot}$  \\
   & ($10^{21}$ g) & ($R_{\male}$) &   & ($^{\circ}$) & ($^{\circ}$) \\
\hline
13  & 360 & 1.77 & 0.13 & 4.2 & 3.3 \\
    & 450 & 2.52 & 0.02 & 5.4 & \\
    & 180 & 4.60 & 0.06 & 7.5 & \\
    & 0.6 & 33.5 & 0.44 & 50.0 & \\
29 & 0.6 & 2.12 & 0.34 & 1.37 & 5.7\\
   & 0.6 & 9.03 & 0.51 & 30.2 & \\
42  & 2.4 & 1.99 & 0.01 & 1.55 & 2.8 \\
    & 5.3 & 2.56 & 0.03 & 0.37  &  \\
    & 0.6 & 2.91 & 0.06 & 1.30  & \\
    & 8.4 & 3.44 & 0.03 & 0.51 & \\
    & 6.6 & 4.45 & 0.02 & 0.31 & \\
    & 10.2 & 5.29 & 0.02 & 0.32 & \\
    & 6.0 & 6.37 & 0.03 & 0.54 & \\
    & 12.6 & 8.36 & 0.05 & 1.22 & \\
HR1 & 3.0 & 5.24 & 0.64 & 7.30 & 9.4 \\
HR9 & 28.8 & 2.30 & 0.03 & 4.30 & 8.6 \\
    & 7.8 & 3.05 & 0.24 & 7.60 & \\
    & 0.6 & 11.8 & 0.31 & 123.1 & \\
\hline
10 & $8 \times 10^{-5}$ & 6.30 & 0.53 & 4.20 & 2.7 \\
  & $10^{-5}$ & 93.9 & 0.59 & 34.1 & \\
26 & $6 \times 10^{-5}$ & 26.6 & 0.21 & 168 & 6.4\\
   & $2 \times 10^{-5}$ & 78.2 & 0.40 & 166 & \\
30 & $7 \times 10^{-5}$ & 34.7 & 0.35 & 147 & 7.3\\
44 & $1.5 \times 10^{-4}$ & 5.80 & 0.43 & 151 & 9.8 \\
   & $3 \times 10^{-5}$ & 65.6 & 0.35 & 155 & \\
\hline
\end{tabular}
\end{minipage}
\end{table}

The end result in each case is a handful of surviving bodies. In cases where there are only
a few surviving particles to start with, the collisional evolution is limited and the
surviving bodies retain substantial eccentricity and inclination (e.g. \# 29 or HR1).
 In the cases
where the collisional evolution starts with many surviving bodies, the effective dissipation resulting from the collisions
reduces the eccentricity and inclinations of the survivors, reproducing systems that bear
some qualitative similarities to the Martian moons (although still somewhat more massive), in
that they damp down to quasi-circular, equatorially aligned orbits. Simulations \# 13, \# 42 and HR9 are representative of this.

The cases simulated here are very simplistic representations of a moon-forming Martian disk, but they
do illustrate the general property that a dynamically excited, reduced mass disk can evolve collisionally
towards a state that shows the same general properties of the Martian moon system.

\subsection{Three Body Capture Orbits}
\label{3BCap}

In the case where the satellite population is acquired during three-body encounters (\S~\ref{Capture}),
the orbits now contain no signatures of a disk, but are rather highly elliptical, isotropically oriented and weakly bound.
 Indeed many have semi-major axes that are substantial fractions
of the Martian Hill sphere and so are potentially susceptible to destabilisation by long term solar perturbations. 
We have also shown (\S~\ref{Demographics}) that the amount of mass likely to be accreted in any particular episode is probably
only a fraction of a Phobos mass.
However, the final satellite mass need not be added in single increments of $10^{18}$g bodies. If a swarm of smaller bodies
is captured in a given encounter, the timescale for collisional evolution is short compared to the Myr timescales
until the next close encounter. We can see this by using the same expression as in \S~\ref{Collision}, where
even bodies at the distance of the Hills sphere should collide on timescales of a few hundred years.

In order to assess the long-term viability of these bound orbits, we once again take the output of our test
particle calculations and evaluate their evolution if we assign mass. We take
 the final positions
and velocities for the captured objects from each calculation in \S~\ref{Capture} as the starting conditions for another integration, using
the {\tt Mercury} Bulirsch-Stoer integrator.  We adopt
a co-ordinate system centered on the TMA and include the Sun as a distant perturber, using the orbital parameters of the
original encounter. We once again include the present-day
value for the Martian  $J_2$. As an example, we take the 
 orbital elements of the 162 objects remaining
 from the last encounter in simulation~26 and 
 integrate for $1.5 \times 10^5$ years (with a 0.01~day timestep) to allow for the collisional evolution to proceed. Each object is assigned
a mass of $10^{16}$g and a density of $2 g/cm^3$. These masses are considerably smaller than those in \S~\ref{Collision}
because we expect a smaller yield due to the considerations in \S~\ref{Demographics}.

The resulting system is again highly collisional, and the rapid collision and accumulation of the captured
planetesimal swarm results in two surviving bodies. Twenty five of the initial bodies are ultimately
ejected by solar perturbations, and the rest collide with the central TMA. Most of this latter population are the
products of collisions between satellites -- when bodies with substantial eccentricites and approximately equal
proportions of prograde and retrograde orbits collide,
 the cancellation of the vector angular
momentum often results in very small periareia. The innermost surviving object has a mass of $ 6 \times 10^{16}$g,
a semi-major axis of 26.6 $R_{\male}$, an eccentricity of 0.21 and an inclination of $168^{\circ}$. The other
survivor has a mass of $2 \times 10^{16}$g, semi-major axis of 78.2 $R_{\male}$, eccentricity of 0.40 and inclination
of $165.5^{\circ}$. This system is still not a very good analogue for the Martian moons, but it does illustrate
that collisional evolution can trap a fraction of the captured bodies into bound orbits and on timescales
short compared to the inverval between large body encounters. A more realistic evaluation of this scenario
would then require simulating the successive injection of new captured objects into the collisional remnants
of prior episodes to see whether the gradual accumulation of small bodies can lead to a viable moon system. 

We have repeated this calculation for each of the cases in \S~\ref{Capture} that produced captured populations
of fifty or more particles. The results are shown in the second group in Table~\ref{GroundDown}. We see that the
character of the surviving satellite swarms is one or two bodies that remain in a quasi-equatorial orbit (although
frequently retrograde!). We find that, in two cases (Simulations \#~10 and \#~44), the collisional evolution
produces a surviving body with a semi-major axis that lies between those of Phobos and Deimos. In addition, all
four simulations produced surviving bodies with semi-major axes in the range 10--100 $R_{\male}$. These are well
outside the present orbit of Deimos and so could possibly represent an as yet undiscovered population of objects.
However, these would likely be unbound by any close passages of massive bodies and also possibly by longer time
perturbations experienced by the Martian system. Indeed, Sheppard, Jewitt \& Kleyna (2004) place stringent limits
(size $< 0.09$km) on any irregular satellites in the Martian system.

\subsection{Tidal Evolution}

We have thus far confined our attention to the gravitational influences felt by the Martian system over the first $10^8$ years
of evolution. Of course, in the interval between then and today, possible moon systems can have evolved substantially
via tides. Therefore, a comparison with present day observations should also account for possible changes to the satellite
system over the intervening period.

Rocky bodies such as Phobos and Mars are believed to exhibit a tidal response that is rather weakly
dependant on frequency, and is conveniently described with a `constant Q' formalism (see Goldreich \& Soter 1966;
Wisdom \& Tian 2015 for discussion). We will adopt the expressions from Jackson, Greenberg \& Barnes (1998), where Mars
is now the central object.

Circularisation of eccentric satellite orbits should occur on a timescale
\begin{eqnarray}
T_{e} & = & 5 \times 10^7 {\rm years} \left( \frac{a}{10 R_{\male}}\right)^{13/2} \frac{Q_1}{100} \nonumber \\
&&  \left( \frac{M}{0.1 M_{\oplus}} \right)^{-3/2}
\left( \frac{R}{10 km} \right)^{-2}
\end{eqnarray}
where $M$ is the mass of the TMA, $R$ is the radius of the satellite (assuming a constant density sphere with density $3 g/cm^3$) and
assuming $Q_1=100$ (Le Maistre et al. 2013). The $Q_1$ value here is for the satellite, which dominates the tidal dissipation that leads to
circularisation. Once the satellite orbit is circularised, the orbit will continue to evolve due to dissipation in the planet. The
satellite spirals in or out, depending on whether it lies interior or exterior to the synchronisation radius.

The timescale for this evolution is 
\begin{eqnarray}
T_{in} &  = & 1.2 \times 10^9 {\rm years} \left( \frac{a}{5 R_{\male}}\right)^{13/2}  \frac{Q_2}{100} \nonumber \\
&& \left( \frac{M}{0.1 M_{\oplus}} \right)^{1/2}
\left( \frac{R}{10 km} \right)^{-3}
\end{eqnarray}
where $Q_2$ now characterizes the dissipation in Mars itself (Black \& Mittal 2015).

We can now apply this to the different surviving asteroid populations described in Table~\ref{GroundDown}.
Perhaps the most interesting is the results for simulation~42 in Table~\ref{GroundDown}. In this case,
 we find that the inner five potential
moons should spiral in within $10^8$ years, but that there are three more with longer evolution times.
The outermost is also outside co-rotation for a Mars-like spin and so represents a good Deimos analogue.
The one at $6.4 R_{\male}$  should spiral in on $\sim 10^9$ year timescales and so represents a good
starting point for a potential Phobos analog -- possibly through an erosional process such as described
by Hesselbrock \& Minton (2016). 
The inner moons of simulation~13 are more massive and should spiral in to Mars in short order, leaving
one distant body largely unaffected by tides. In the case of simulation~29, the innermost body spirals
in quickly, but the outermost may evolve to a Deimos-like orbit. Similarly, the innermost body of HR1
should spiral in within a Myr.

 Of course, much of this rapid evolution is a function of the larger mass
of these bodies, which is set by our simulation limits. Given that the cross-sections for collision
are geometric at these masses, we expect the evolution to be qualitatively similar at lower masses,
which would preserve the configurations shown in Table~\ref{GroundDown} for longer.

A demonstration of this comes from applying similar considerations to the captured body population
in Table~\ref{GroundDown}. The bodies with semi-major axis $> 10 R_{\male}$ remain largely
unaffected by tides, but the two bodies on more compact orbits (from simulations \#~10 \& \#~44)
are both circularised within 100 Myr, but have inspiral times $> 10^{10}$years. They both end
up with circularised orbits intermediate between Phobos and Deimos (4.5 and 4.7 $R_{\male}$ respectively).

\section{Discussion}

Within the context of an annular origin model for the formation of the terrestrial planets, the dynamical
history of Mars is qualitatively different from that of Earth, and we have shown above that this
has potentially important consequences for the origin of the moon systems in the two cases. If the
Earth's moon did indeed form from debris placed into orbit by an impact with a planetary embryo, the
dynamical histories simulated here (and in prior investigations of this model such as Hansen~2009 or Brasser~2013)
are consistent with this being the result of the last major impact experienced by the planet, as 
measured by cosmochemical measures such as the Hf-W ratio (e.g. Jacobson et al. 2014). The model analogues to
Mars in these simulations, on the other hand, experience their last collision over a wide range of
ages. While some can experience a last collision as late as 790~Myr, most have their last collision at
a much younger age, often within 10~Myr. This is also consistent with the cosmochemical analyses of
Mars (Nimmo \& Kleine 2007; Dauphas \& Pourmand 2011).

The Martian analogues that finish their mass growth on timescales $\sim 10$~Myr continue to experience
significant gravitational perturbations from passing bodies until the inner solar system is completely
cleared of remaining embryos, which happens on a timescale more consistent with the Earth's moon forming age.
Therefore, if we postulate a similar origin for the Martian moons as for Earth's moon, the disk of 
debris from which the Moon forms is subject to substantial perturbations for $\sim 100$Myr. This offers
a potential mechanism by which the mass available for Moon formation can be significantly reduced.

\begin{figure}
\includegraphics[width=84mm]{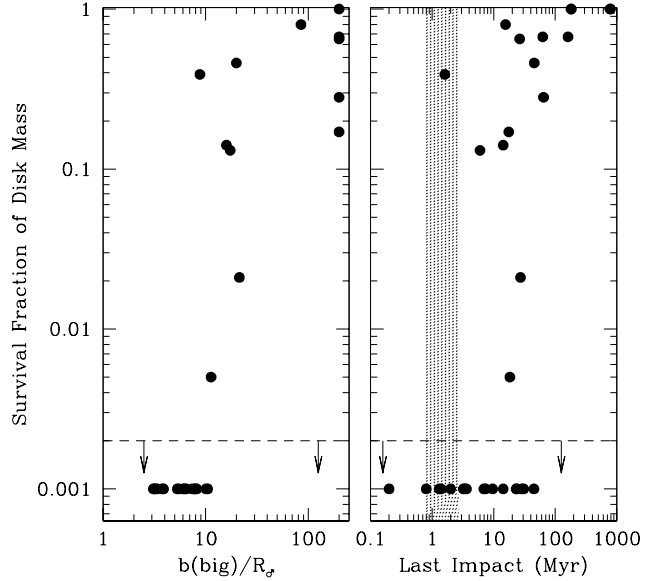}
\caption[Fcomp.ps]{The left-hand panel shows the surviving disk fraction for each of the TMA 
cases (either T or P), as a function of the minimum impact parameter with respect to one of
the surviving terrestrial planets in the same simulation. Only encounters that occur after the
last giant impact on the TMA are counted. The right panel shows the same disk fraction but
now as a function of the age of the last giant impact for the TMA. The shaded region indicates
the claimed age of the last Martian impact. The horizontal dashed lines indicate the minimum
fraction resolvable with our 500 particle starting conditions. Systems with no surviving bodies
are shown below this line.
\label{Fcomp}}
\end{figure}

We have simulated the susceptibility of disk material to the perturbation history of each Martian
analogue found in our simulations. The results are shown in Figure~\ref{Fcomp}. A  wide range of
outcomes is possible.
 We find that  many  systems are completely destroyed
by a sequence of close passages, the most damaging being those which bring the TMA close to one of
the larger forming terrestrial planets. In a few cases, when a TMA had its last collision late or sufficiently far
from the original annulus, it was able to avoid any close passages with massive bodies, and a substantial
fraction of the disk survived, albeit in a dynamically excited state. Most intriguingly, a few TMA
lost nearly all their disk, but did retain a small fraction of the mass. This may offer the most promising
pathway to forming the observed, low mass, Martian moons. Although a wide range is 
theoretically possible, the cosmochemical constraints on the formation of the  Martian system put the
observed age at the low end of the possible distribution and so 
 suggests that any Mars protolunar disk experienced substantial perturbation after its formation.

Craddock (2011) has estimated the size of the putative impactor based on both the spin of the planet and
the size of the largest observed craters. Within the framework of spin being determined by stochastic
accretion (Dones \& Tremaine 1993), he estimates a mass $\sim 10^{25} g \sim 0.0017 M_{\oplus}$, which
is of similar size to the initial embryo masses in our simulation. Similar estimates are obtained by
applying emprirical scaling laws to the largest Martian basins such as Borealis or Elysium. Thus, our
simulations resolve the mass scale invoked for the disk-forming impactors. The principal uncertainty that follows from
such a scenario is the size of the resulting disk. Simulations of the formation of possible protolunar
disk (e.g. Citron, Gendra \& Ida 2015; Canup \& Salmon 2016; Hyodo et al. 2017) generically produce moons that are too
large. This can potentially be mitigated depending on where the mass the deposited relative to the Roche
limit (Rosenblatt \& Charnoz 2012) or if there is substantial evolution in the Moon system that leaves
only a fraction of the original mass visible today (e.g. Rosenblatt et al. 2016; Hesselbrock \& Minton 2016).
Our results suggest that such evolution can be the consequence of an extended series of external perturbations.

Another implication of this is that the perturbations heat the satellite system dynamically, so that the 
 histories for the moon systems described here are substantially collisional. As a result, some of our calculations above, which
utilise perfect merging upon collision, are probably naive. On the other hand, the low densities of Phobos and Deimos (Andert et al. 2010;
P\"{a}tzold et al. 2014) 
are often used as arguments to suggest that they are rubble piles, likely substantially fractured during prior evolutionary
stages. This is eminently consistent with the energetics of the above scenarios. Mutual collision velocities in
a dynamically perturbed  disk  are expected to be $\sim \sqrt{e} V_{orb}$ on average. At distances of $3 R_{\male}$ from
Mars, orbital velocities are $\sim 2$km/s, suggesting that, for $e \sim 0.5$, the kinetic energy of impact is $\sim 2 \times 10^{29}$ergs
for Phobos-sized bodies. For solid bodies, assuming a chemical bond energy $\sim 0.1$eV per $m_p$, the binding energy of an
equivalent body is $\sim 6 \times 10^{29}$ergs. Thus, we expect collisions to have a substantial, but not necessarily catastrophic,
effect on the survivors.

\begin{figure}
\includegraphics[width=84mm]{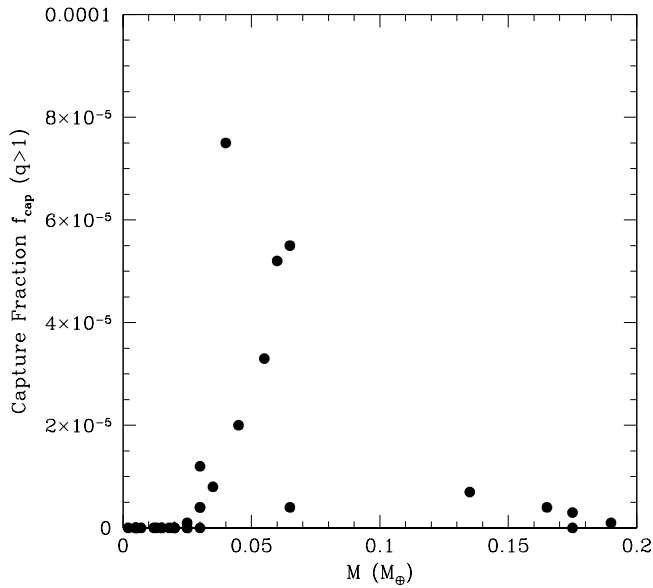}
\caption[CapM.ps]{Each point shows the empirical capture rate during the last close passage, after accounting for the
fraction lost by impacts on to the surface. This is shown as a function of planet mass, and it clearly favours masses
$\sim 0.05 M_{\oplus}$.
\label{CapM}}
\end{figure}

Given the possibility that moon forming disks may be completely destroyed, we
 have also considered the possibility that small bodies could be captured during three-body encounters. 
We find that it is indeed possible to produce a captured
population of bodies, but that the amount of mass captured is likely to be less than a Phobos amount per interaction
(if normalised by the Late Veneer mass estimates for Mars). Such a population is also very weakly bound, more like
the irregular satellites of the giant planets than the observed Martian system. However, such a swarm of small bodies
would evolve rapidly by collisions and our estimates in \S~\ref{3BCap} suggest that it may be possible to produce Deimos-like objects
via a sequence of collisions. The capture process also induces an order of magnitude more mass to strike the Martian
surface. If some of these impacts generate orbital debris, that may also help to dissipate the orbital energy and
damp the eccentricities and inclinations of the captured objects. The efficiency of this capture
process is also a strong function of TMA mass, as shown in Figure~\ref{CapM}. For planets with mass $> 0.1 M_{\oplus}$ the
efficiency of capture is quite small, and for masses $<0.03 M_{\oplus}$ the capture process is more efficient, but nearly
all the objects strike the surface of the planet. Only in the region $0.03 M_{\oplus} < M < 0.07 M_{\oplus}$ do we see
the creation of a population of bound bodies with the potential for longer lifetimes. The mass estimate could also
be increased if there is a substantial reservoir of small bodies that survive to late times (e.g. O'Brien, Morbidelli
\& Levison 2006; Jacobson et al. 2014) or if some of the orbiting debris is actually planetary material dislodged
in a previous collision (Kokubo \& Genda 2010; Chambers 2013).

The different potential pathways have potentially different cosmochemical implications. If the Martian moons are
simply the dynamically eroded remnants of an originally larger population, then their cosmochemical signatures
should represent the composition of Mars (with a possible contribution from the impactor). Although this would
retain all the current issues arising from spectral comparisons, we do note that the increased collisional erosion
may help to provide the micron-sized dust required by Ronnet et al. (2016). In the event that the moons are the
consequence of a three-body capture scenario, the possibilities are broader. The source population for captured
bodies is the same as for the original planetary embryos, and so the signatures should still be representative
of the inner solar system. In classical assembly models, where each planet assembled from material that condensed
from the nebula at that location, chemical signatures are held to be representative of the distance at which the
material sedimented. However, an annular model like the one invoked here is likely the result of a far more
dynamic history, either via the Grand Tack (Walsh et al. 2011) or a planet trap in the original disk (Hansen 2009).
In such an event, the annulus may contain material from a wide range of chemical compositions.
Historically, capture models for the Martian moons have adopted the Asteroid belt as the original source population.
 This is potentially only a semantic difference, as it has been suggested that the
 S-type asteroids are a population scattered out of the same birth annulus whose orbits were damped by gas drag
(Raymond \& Izidoro 2017). 

A further question is the final disposition of material lost from the original satellite population, and where
it might reside at present. Most of the material presumably re-impacts the surface of Mars on a hyperbolic
orbit at some future date, but this scenario may also provide a pathway to emplace objects at the Martian trojan
points. The Martian Trojans show spectral features of Olivene that suggest
a Martian surface origin (Rivkin et al. 2003; Scholl, Marzari \& Tricarico 2005;  Borisov et al. 2017; Polishook et al. 2017). Furthermore, the
 Eureka family of Martian Trojans appears to have an 
age $> 1$~Gyr (Scholl et al. 2005; Christou 2013; Cuk et al. 2015,
de la Fuente Marcos \& de la Fuente Marcos 2013) which points to a long-lived, possibly almost primordial
population. Polishook et al. (2017) propose that such material results directly from Martian surface impacts drawn
from the Grand Tack Scenario (Walsh et al. 2011). The Martian dynamical histories in that model and ours are very similar, and
the Trojans could result either as envisioned by Polishook or from the material that condensed in a Moon forming disk
that was then dispersed by subsequent perturbations, as described here.
 One datum potentially at odds with this scenario is 
the lack of similar spectral features in Phobos \& Deimos, but it could be consistent if the lack of a feature is
because of the presence of fine-grained dust generated by a more extended collision history (e.g. Guiranna et al. 2011;
Ronnet et al. 2016).

\section{Conclusion}

The scenario in which the small mass of Mars relative to Earth is due to its outward
migration from a common formation zone also offers a pathway to explaining the small
mass of the Martian moons. If the original moon forming material was produced in a 
giant impact, the nascent moon system would have been subject to an extended period
of gravitational perturbations during the scattering-induced migration. 

We have presented preliminary calculations of the consequences of these perturbations,
both for the survival of a pre-existing disk and for the capture of additional material.
In both cases, the end products of the evolution can reproduce some of the basic properties
of the moons of Mars, despite the relatively crude numerical resolution. Future, more detailed
calculations will provide more detailed information on these questions.

\section*{Acknowledgements}

The author thanks the referee, Apostolos Christou, for a constructive referee report.
 The simulations described here were performed
on the UCLA Hoffman2 Shared computing cluster and using the resources provided
by the Bhaumik Institute.
This research has made use of NASA's Astrophysics Data System.

\end{document}